\documentclass[a4paper,10pt,preprint,twocolumn,3p]{elsarticle}

\usepackage{amsthm}
\usepackage{amssymb}
\usepackage{mathtools}
\usepackage{physics}
\usepackage{graphicx}
\usepackage{adjustbox}
\usepackage{placeins}
\usepackage[T1]{fontenc}
\usepackage{multirow}
\usepackage{dcolumn}
\usepackage{bm}
\usepackage{color}
\usepackage{comment}
\usepackage{booktabs}
\usepackage[dvipsnames]{xcolor}
\usepackage{enumitem}

\definecolor{nndcblue}{RGB}{111, 201, 224}
\definecolor{nndcpink}{RGB}{231, 140, 198}
\definecolor{linkblue}{RGB}{0, 128, 172}
\usepackage[colorlinks = true,citecolor=nndcpink,urlcolor=nndcpink,linkcolor=nndcpink]{hyperref}

\usepackage{orcidlink}

\journal{Physics Letters B}

\biboptions{numbers,sort&compress}
\begin{document}

\title{Two-neutrino double-weak decays of $^{126}$Xe and $^{134}$Xe\\ from different many-body methods}

\author[add1,add2,add3]{C. Brase\,\orcidlink{0000-0002-5876-7621}}
\ead{catharina.brase@tu-darmstadt.de}
\author[add1,add2]{L. Jokiniemi\,\orcidlink{0000-0002-9327-5868}}
\ead{lotta.jokiniemi@tu-darmstadt.de}
\author[add4,add5]{E. Kauppinen\,\orcidlink{0000-0003-3517-4978}}
\ead{elina.k.kauppinen@jyu.fi}
\author[add6]{B. Romeo\,\orcidlink{0000-0001-8896-4565}}
\ead{bromeo@unc.edu}
\author[add5,add7]{J. Kotila\,\orcidlink{0000-0001-9207-5824}}
\ead{jenni.kotila@jyu.fi}
\author[add8,add9]{J. Men\'{e}ndez\,\orcidlink{0000-0002-1355-4147}}
\ead{menendez@fqa.ub.edu} 
\author[add1,add2,add3]{A.~Schwenk\,\orcidlink{0000-0001-8027-4076}}
\ead{schwenk@physik.tu-darmstadt.de}
\address[add1]{Technische Universit\"at Darmstadt, Department of Physics, 64289 Darmstadt, Germany}
\address[add2]{ExtreMe Matter Institute EMMI, GSI Helmholtzzentrum f\"ur Schwerionenforschung GmbH, 64291 Darmstadt, Germany}
\address[add3]{Max-Planck-Institut f\"ur Kernphysik, Saupfercheckweg 1, 69117 Heidelberg, Germany}
\address[add4]{Department of Physics, University of Jyväskylä, P.O. Box 35, Jyväskylä FI-40014, Finland}
\address[add5]{International Center for Advanced Training and Research in Physics (CIFRA), 409, Atomistilor Street, Bucharest-Magurele, 077125, Romania}
\address[add6]{Department of Physics and Astronomy at UNC, 27999 Chapel Hill, USA}
\address[add7]{Finnish Institute for Educational Research, University of Jyväskylä, P.O. Box 35, Jyväskylä FI-40014, Finland}
\address[add8]{Departament de Física Quàntica i Astrofísica, Universitat de Barcelona, 08028 Barcelona, Spain}
\address[add9]{Institut de Ci\`encies del Cosmos, Universitat de Barcelona, 08028 Barcelona, Spain}

\begin{abstract}
We calculate the nuclear matrix elements and corresponding half-lives for the two-neutrino double-electron capture of $^{126}$Xe and the two-neutrino double-beta decay of $^{134}$Xe. We use different many-body methods: the proton-neutron quasiparticle random-phase approximation, the nuclear shell model, the microscopic interacting boson model, and an effective field theory for heavy nuclei. For both nuclei, all our half-life predictions are generally consistent with each other when including theoretical uncertainties for each method. Interestingly, for all calculations the lower range of the predicted $^{134}$Xe half-life is shorter than $T^{2\nu}_{1/2} \approx 2\times10^{24}$\,y, which may be within the reach of next-generation experiments. For $^{126}$Xe, our results typically predict one order of magnitude longer half-lives than those for $^{134}$Xe. 
\end{abstract}

\maketitle

\section{Introduction}

Double-beta ($\beta\beta$) decay and double electron capture (ECEC) offer some of the most promising avenues to test fundamental symmetries beyond the standard model of particle physics. In $\beta\beta$ decay, two neutrons become two protons, while the reverse occurs in ECEC; in $\beta\beta$ decay, two electrons are emitted, whereas in ECEC they are captured. If either of the two processes occurs without neutrino emission, the number of leptons is not conserved, breaking a standard model symmetry~\cite{Agostini:2022zub,Dolinski:2019nrj,Gomez-Cadenas:2023vca}. Because of the deep consequences this would have, several international collaborations are pursuing a neutrinoless $\beta\beta$ decay or ECEC detection~\cite{Agostini:2022zub,Dolinski:2019nrj,Gomez-Cadenas:2023vca}.

While the neutrinoless decay is their main goal, these experiments are also sensitive to two-neutrino ($2\nu$) $\beta\beta$ decay and ECEC~\cite{Saakyan:2013yna}. These modes are permitted by the standard model of particle physics, even though they are extremely rare and represent the longest half-lives ever observed in the laboratory~\cite{Barabash2020}. 

The study of $2\nu$ double-weak processes is important in their own right. Since neutrinoless and $2\nu$ modes share the same initial and final nuclear states, $2\nu$ measurements can provide insights into the sought-after neutrinoless counterpart. This is especially important for nuclear matrix elements  (NMEs), which are poorly known for neutrinoless decays~\cite{Engel2017}. Indeed, various nuclear many-body approaches have observed a correlation between $2\nu$ and neutrinoless NMEs~\cite{Jokiniemi:2022ayc,Horoi:2022ley,Horoi:2023uah,Horoi:2026fxp,Lian:2026avy}. Therefore, $2\nu$ double-weak decays provide valuable information for the many-body methods used to predict neutrinoless NMEs.

About a dozen $2\nu$  double-weak decays have been measured so far~\cite{Barabash2020}, including transitions to both ground and excited states. While in most cases theoretical calculations have not been able to predict the half-lives prior to their measurement, a remarkable success was the case of the $^{124}$Xe ECEC~\cite{XENON:2019dti,XENON:2022evz,LZ:2024wvs,PandaX-4T:2024fls}, measured in good agreement with the predictions of three different many-body methods~\cite{CoelloPerez:2018ghg,Pirinen2015,Suhonen:2013rca}. This supported the theory input obtained with the same approaches for the neutrinoless $\beta\beta$ decay of $^{136}$Xe, one of the isotopes used in some of the leading neutrinoless searches~\cite{KamLAND-Zen:2024eml,PandaX-4T:2025jel,nEXO:2021ujk,NEXT:2020amj,XLZD:2024pdv}. 

In this Letter, we predict the $2\nu$ half-life of two additional otherwise stable xenon isotopes: $^{126}$Xe, with ECEC $Q$-value $Q_{\rm ECEC}=917.8(14)\text{ keV}$ and natural abundance of $0.089(3)\%$, and $^{134}$Xe, with $Q_{\beta\beta}=824.2(3)\text{ keV}$ and abundance of $10.436(35)\%$~\cite{nudat3}. Neither of these decays have been detected yet~\cite{Akerib2020,XMASScoll18,EXO2002026,Yan24} but current and future experiments using xenon can enhance the sensitivity to the corresponding half-lives, however a $^{126}$Xe ECEC signal will be very similar to a $^{124}$Xe ECEC one. As in a previous study on the $2\nu\beta\beta$ decay of $^{136}$Xe into an excited state in the final nucleus~\cite{Jokiniemi:2022yfr}, we use four  different widely-used many-body methods: the proton-neutron quasiparticle random-phase approximation (pnQRPA)~\cite{Vogel:1986nj,Engel:1988au,Ejiri:2019ezh}, the nuclear shell model (NSM)~\cite{Caurier:1990dc,Caurier12,Horoi13,Coraggio:2023eep}, the microscopic interacting boson model (IBM-2)~\cite{Barea:2015kwa}, and an effective field theory (EFT) for heavy nuclei~\cite{CoelloPerez_2025,CoelloPerez_2018}. 
Previous calculations of $2\nu$ double-weak decays of $^{126}$Xe and $^{134}$Xe 
include different versions of the pnQRPA~\cite{Pirinen2015,Delion:2017bie,Raduta:2005be} and IBM-2~\cite{Barea:2015kwa}, as well as projected Hartree-Fock-Bogoliubov (PHFB) calculations~\cite{Singh:2007jh,Shukla07}.

In our work, we place emphasis in estimating the theoretical uncertainties for each many-body approach to provide more reliable ranges. For our phenomenological pnQRPA, NSM, and IBM-2 results, we take into account, based on systematics within each method, the uncertainty from quenching, i.e., the NME overestimation due to missing two-body currents and many-body correlations~\cite{Gysbers2019a,Engel2017}. These effects can be included in ab initio approaches~\cite{Gysbers2019a,Stroberg:2021guc,Li:2025exk} with promising developments on $2\nu$ double-beta decays~\cite{Novario2021,Lian:2026avy,Li2026}. Likewise, in the EFT for heavy nuclei these effects are captured by construction and uncertainties can be estimated systematically from higher orders~\cite{CoelloPerez_2025,CoelloPerez_2018}.   

\begin{figure}[t!]
    \centering
    \includegraphics[width=\linewidth]{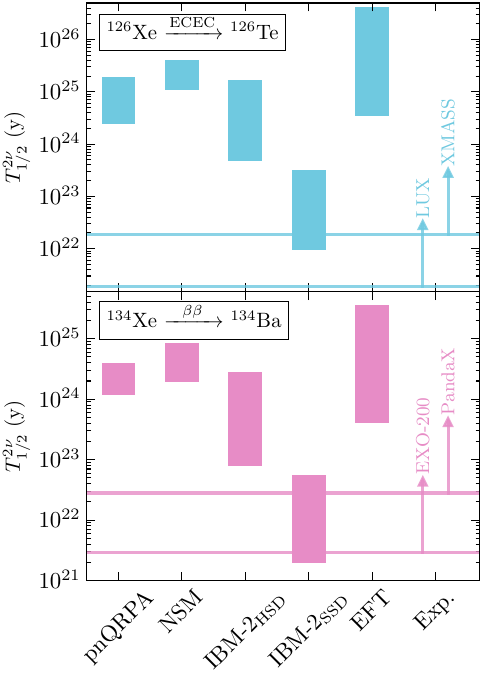}
    \caption{Ground-state-to-ground-state half-life predictions (listed in Table~\ref{tab:prediction_summary}) for $2\nu$ECEC of $^{126}$Xe (upper panel) and $2\nu\beta\beta$ decay of $^{134}$Xe (lower panel) for the different many-body methods, compared to the latest experimental limits (Exp.) from LUX~\cite{Akerib2020}, XMASS~\cite{XMASScoll18}, EXO-200~\cite{EXO2002026} and PandaX~\cite{Yan24}. Note that the theoretical uncertainties do not have a statistical meaning and are obtained differently in each approach.
    }
    \label{fig:T12_allmtheods}
\end{figure}

Figure~\ref{fig:T12_allmtheods} summarizes our half-life predictions, which are consistent across all many-body methods used. They are compared to experimental lower limits~\cite{Akerib2020,XMASScoll18,EXO2002026,Yan24}, which only disfavor part of the IBM-2 results. For $^{134}$Xe, the lower part of our predictions for all methods could be within the reach of next-generation experiments~\cite{LUXZEPLIN21}. 

\section{Nuclear matrix element}
\label{sec:nme}

The half-life for $2\nu\beta\beta$- and $2\nu$ECEC-decay can be written as the product of the phase-space factor (PSF) $G^{2\nu}$ and the NME $M^{2\nu}$:
\begin{align}
\big(T^{2\nu}_{1/2}\big)^{-1} =  G^{2\nu} g_\mathrm{A}^4 (M^{2\nu})^2\,,
\label{eq:half-life}
\end{align}
where $g_\mathrm{A} \approx 1.27$ denotes the axial-vector coupling. The PSF corresponds to the lepton contribution and the NME to the nuclear one, which is the main focus of this work. Corrections to Eq.~\eqref{eq:half-life} take into account a Taylor expansion of the energy denominator of the NME~\cite{Simkovic:2018rdz}, as well as subleading terms~\cite{Morabit:2024sms}. We neglect these contributions, since they are typically at the few-to-ten percent level~\cite{Simkovic:2018rdz,Morabit:2024sms,Castillo2025}, which is much smaller than the theoretical uncertainties considered in our work. For more information on the PSF, see \ref{app:all} as well as Refs.~\cite{Kotila2012,Kotila:2013gea}. 

The $2\nu$ NME is given by~\cite{Faessler1998,Elliott2002,Engel2017}
\begin{align}
\label{eq:M2nu}
    M^{2\nu}=&\sum_{k}\frac{\langle0^+_{ f}||\sum_a\tau^-_a\boldsymbol{\sigma}_a||1^+_k\rangle\langle1^+_k ||\sum_b\tau^-_b\boldsymbol{\sigma}_b||0^+_{i}\rangle}{(E_k-(E_i+E_f)/2)/m_e}\,,
\end{align}
where $|0^+_{i}\rangle$ and $|0^+_{f}\rangle$ denote the ground states of the initial and final nuclei, respectively. The sum over $k$ runs over all intermediate $1^+$ states and $a$ and $b$ over all nucleons. The operators $\tau^-$ and $\boldsymbol{\sigma}$ represent isospin lowering and spin, respectively. The denominator contains the energies $E$ of the initial $i$, final $f$ and $k$th intermediate $1^+_k$ state and is divided by the electron mass $m_e$ for a dimensionless NME. In all calculations, we adjust the energy denominator for the lowest $1^+_1$ state to the experimental value. We summarize the values of the PSFs, ground-state and intermediate-nucleus $1^+_1$ energies in Table~\ref{tab:quantities_we_all_use}. 

In addition, we define an effective NME~\cite{Barabash2020,Kotila2012}
\begin{equation}
M^{2\nu}_{\rm{eff}}=q^2g_{\rm A}^2\,M^{2\nu}\,,
\label{eq:effective_NME}
\end{equation}
which also includes the quenching factor, $q<1$, required for pnQRPA, NSM and IBM-2 calculations to reproduce experimental data. 

\section{Many-body methods}
\label{sec:many-body_methods}

\subsection{Proton-neutron quasiparticle random-phase approximation}
\label{subsec:pnQRPA}

In the spherical pnQRPA method, the $0^+$ ground states of the initial and final nuclei are taken as QRPA vacua, $\ket{\rm QRPA}$, and states in the intermediate odd-odd nuclei are built on top of them~\cite{Suhonen2007}. The $1^+_k$ states in the intermediate nucleus, appearing in the expression of the NME, Eq.~\eqref{eq:M2nu}, are obtained by performing pnQRPA diagonalizations based on the initial and final states as
\begin{align}
    \ket{1_k^{+}}=
    \sum_{pn}\left(X^{1_k^{+}}_{pn}[a^{\dagger}_pa^{\dagger}_n]_{1}-Y^{1_k^{+}}_{pn}[a^{\dagger}_pa^{\dagger}_n]^{\dagger}_{1}\right)\ket{\rm QRPA}\,,
\end{align}
where $a_p^{\dag}$ ($a_n^{\dag}$) creates a proton (neutron) quasiparticle in the orbital $p$ ($n$), and $X$ and $Y$ are the backward and forward pnQRPA amplitudes. Here we omit projection quantum numbers $M$ for simplicity.

There are a number of parameters that need to be adjusted in the pnQRPA: the particle-hole parameter $g_{\rm ph}$ is fitted to reproduce the semi-empirical energies of the Gamow-Teller (GT) giant resonances in the intermediate nuclei $^{126}$I and $^{134}$Cs~\cite{Suhonen2007}; and the particle-particle parameter $g_{\rm pp}$ is fitted to the $2\nu$ half-lives of the neighboring double-weak-decaying isotopes $^{124}$Xe and $^{136}$Xe, following the partial isospin-restoration scheme~\cite{Simkovic2013} (for more details see \ref{app:pnQRPA}). We assume the range $g_{\rm A}^{\rm eff}=qg_{\rm A}=(0.8-1.27)$ of the effective axial coupling, corresponding to a quenching factor $q=(0.63-1.0)$. The energies of the intermediate $1^+_k$ states are shifted so that the lowest state matches experiment. Since the excitation energy of the lowest $1^+_1$ state in $^{124}$I is not experimentally known, we use a conservative range of $E^{\rm exc}(^{124}{\rm I},1^+_1)=50-300$\,keV, based on neighboring iodine isotopes, in the fitting procedure of the parameter $g_{\rm pp}$ for $^{126}$Xe. 

We consider large no-core single-particle bases consisting of the lowest 25 single-particle orbitals~\cite{Jokiniemi2018,Jokiniemi2021} in a Woods-Saxon potential. The proton and neutron quasiparticle spectra, needed in the pnQRPA diagonalization, are obtained by solving the BCS equations with the two-body interaction derived from the Bonn A one-boson-exchange potential~\cite{Holinde1981} fine-tuned by adjusting the proton and neutron pairing parameters to pairing gaps extracted from proton and neutron separation energies~\cite{Suhonen2007}.

The current study improves the results presented in Ref.~\cite{Pirinen2015}, which uses smaller single-particle bases and a different parameter fitting procedure based on single-$\beta$ decays without isospin restoration. On the other hand, our method assumes spherical nuclei, whereas the small deformation effects of the involved nuclei have been taken into account in previous studies based on different versions of deformed pnQRPA~\cite{Raduta:2005be,Delion:2017bie}.

\subsection{Nuclear shell model}
\label{subsec:NSM}

The NSM, also known as the configuration interaction method, solves the nuclear Schrödinger equation for medium- and heavy-mass nuclei by separating an active valence space, where the dynamics of nucleons are explicitly treated, from a core of fully occupied orbitals. This framework successfully describes a wide range of properties of light- and medium-mass nuclei~\cite{MPinedo,Brown01,Otsuka19}, including GT $\beta$-decays~\cite{Wildenthal85,Chou93,Yoshida18,Coraggio20}.

As in previous studies~\cite{Caurier12,Horoi13,Nitescu:2024ppf}, we use the configuration space comprising the single-particle orbitals $1d_{5/2}$, $0g_{7/2}$, $2s_{1/2}$, $1d_{3/2}$, and $0h_{11/2}$ both for protons and neutrons with the effective GCN5082~\cite{Caurier:2010az} and QX~\cite{QiQX} interactions. Note that GCN5082 describes the low-lying nuclear structure of $^{136}$Cs, the intermediate nucleus in the $^{136}$Xe $2\nu\beta\beta$ decay, significantly better~\cite{Rebeiro:2023kvs}. This indicates a possible deficiency of the QX interaction which may extend to $^{134}$Cs, the intermediate nucleus of the $^{134}$Xe $2\nu\beta\beta$ decay. While NSM calculations exist for the decay of $^{124}$Xe and $^{136}$Xe (isotopes neighboring those of this work), here we report the first NSM study of the $2\nu$ECEC of $^{126}$Xe and $2\nu\beta\beta$ of $^{134}$Xe~\cite{Nitescu:2024ppf,Caurier12,Horoi13}.

We obtain the NMEs by summing over intermediate states using the Lanczos method, as described in Ref.~\cite{MPinedo}. We increase the number of intermediate states until reaching convergence, which typically implies few tens of iterations. In the case of the $2\nu$ECEC of $^{126}$Xe an additional approximation is needed, allowing only configurations up to three nucleons excited from the lower-energy orbitals $1d_{5/2}$, $0g_{7/2}$ to the higher-energy orbitals $2s_{1/2}$, $1d_{3/2}$, and $0h_{11/2}$. For additional details, we refer the reader to~\ref{app:NSM}.

In addition, we apply a quenching factor $q$ to the NSM NMEs, inferred from comparing NSM calculations~\cite{Horoi16,CoelloPerez:2018ghg,Nitescu:2024ppf,Castillo2025} with measured $2\nu\beta\beta$ decays of $^{136}$Xe and $^{130}$Te~\cite{Barabash2020} and from the $2\nu$ECEC of $^{124}$Xe~\cite{XENON:2019dti,XENON:2022evz,LZ:2024wvs,PandaX-4T:2024fls}. For the GCN5082 interaction, the quenching factor is also inferred from a systematic comparison of calculated and experimental GT $\beta$ decays of nuclei in the xenon region~\cite{Caurier12}. This analysis yields the ranges $q=0.42-0.57$ for the GCN5082 interaction and $q=0.67-0.82$ for the QX interaction. The range associated with the quenching factor is the dominant source of NSM NME theoretical uncertainty.

\subsection{Microscopic interacting boson model}
\label{subsec:IBM}

The IBM proton--neutron version,  IBM-2~\cite{ARIMA1977205,iac87}, maps the fermionic nuclear Hamiltonian onto a boson space~\cite{OTSUKA19781} and evaluates NMEs using bosonic wave functions that reproduce important nuclear properties. The method and its application to double-weak decay have been discussed in detail in previous studies~\cite{Barea:2009zza,Barea:2015kwa}.

The IBM-2 Hamiltonian parameters for the initial $^{126, 134}$Xe and final $^{126}$Te, $^{134}$Ba nuclei are determined by fitting their low-lying spectroscopic properties. We adopt IBM-2 parameter sets from previous studies in this mass region when available. For $^{126}$Xe and $^{134}$Xe, we use parameters from Ref.~\cite{GADE2000268}, and for $^{134}$Ba, from Ref.~\cite{PUDDU1980109}. The parameters for $^{126}$Te are determined in the present work (see Table~\ref{tab:IBM2-parameters}). The IBM-2 configuration space is the same as the NSM one. The mapping from fermions to bosons incorporates effective interaction strengths and single-particle energies chosen to yield single-particle occupancies consistent with available nucleon-removal data~\cite{Kotila:2016pib}.

The IBM-2 calculation of the $2\nu$ double-weak NMEs is performed using the closure approximation, in which the energies of the intermediate $1^+_k$ states are replaced by an average excitation energy $\langle E_k\rangle$, and the sum over intermediate states is then carried out analytically. With this approximation, the NME can be written as
\begin{equation}
    M^{2\nu}_{\rm IBM}=\frac{\langle 0^+_{f}||\sum_{a,b}\tau^-_a \tau^-_{b}\,
    \boldsymbol{\sigma}_a\!\cdot\!\boldsymbol{\sigma}_{b}
    ||0^+_{i}\rangle}
    {\left(\langle E_k\rangle-(E_i+E_f)/2\right)/m_e}\,.
    \label{eq:M2nu_IBM2}
\end{equation}
In case of higher-state dominance (HSD), the closure energy $\langle E_k\rangle$ is taken to be the average excitation energy of the intermediate states in the odd-odd $^{126}$I and $^{134}$Cs nuclei, estimated by the systematics as $1.12A^{1/2}$\,MeV. In case of single-state dominance (SSD)~\cite{Abad1984a,Abad1984b,Domin2005} $\langle E_k\rangle$, is simply replaced by the excitation energy of the lowest $1^+_1$ state. For both of the studied decays, we consider SSD and HSD. Our calculations extend the results in a previous IBM-2 study for $^{134}$Xe~\cite{Barea:2015kwa}, which reports only the general closure-approximation result and does not provide separate predictions for the SSD and HSD assumptions.
 
Within the IBM-2 framework, the overestimation of the $2\nu$ double-weak NME arises primarily from omitting important correlations and non-nucleonic degrees of freedom, and is further affected by the use of the closure approximation. To account for these effects, we renormalize the operator by introducing a phenomenological quenching factor~$q$. Following earlier IBM-2 studies~\cite{PhysRevC.87.014315} that reproduce the experimental $2\nu\beta\beta$-decay systematics in this mass region, we consider three possible values: (i) the bare $g_{\rm A}$, $q=1$; (ii) a moderate quenching, $q=0.788$; and (iii) the maximum quenching for IBM-2, $q = A^{-0.18}$ (see Table~\ref{tab:IBM2-NMEs}).  

\subsection{Effective field theory for heavy nuclei}
\label{subsec:EFT}

The EFT for heavy nuclei is formulated in terms of nucleon and phonon degrees of freedom coupled to an even-even spherical core. Its power counting enables systematic quantification of uncertainties, and its analytical form makes the calculation of heavy nuclei computationally straightforward. In addition, the EFT does not require quenching because low-energy constants (LECs) encode unresolved higher-energy physics (hence ${q=1}$)~\cite{CoelloPerez_2018}. For a detailed review of the EFT for heavy nuclei, see Ref.~\cite{CoelloPerez_2025}. 

Following Refs.~\cite{CoelloPerez_2018, CoelloPerez:2018ghg}, we predict $M^{2\nu}$ within the SSD approximation,
\begin{multline}
M^{2\nu}(0^+_{i} \to 0^+_{f})\\
\approx \frac{M_{\mathrm{GT}}(0^+_{i} \to 1_1^+)\,M_{\mathrm{GT}}(1_1^+ \to 0^+_{f})}
{D_{10^+_f}/m_e}\,,
\end{multline}
and quantify the relative truncation uncertainty $\delta_{\mathrm{EFT, gs\rightarrow gs}}^{2\nu}$ of this approximation within the EFT for ground-state to ground-state transitions as~\cite{CoelloPerez_2018}
\begin{multline}
\delta_{\mathrm{EFT, gs\rightarrow gs}}^{2\nu}(0^+_{i} \to 0^+_{f})\\
= \frac{D_{10^+_{f}}}{\Lambda}
\,
\Phi\!\left(
\frac{\omega}{\Lambda},\, 1,\, \frac{D_{10^+_{f}}+\omega}{\omega}
\right)\,,
\end{multline}
where $D_{10^+_{f}}$ corresponds to the energy denominator as in Eq.~(\ref{eq:M2nu}) for the lowest $1^+_1$ energy, $\Lambda$ is the breakdown scale, $\Phi$ the Lerch Transcendent given by $\Phi(z,s,a) = \sum_{n=0}^\infty \frac{z^n}{(n+\alpha)^s}$ and $\omega$ the phonon excitation energy. The latter is determined by the average of the first $2^+_1$ excitation energies in the initial and final nucleus. 

The two LECs in $ M_{\mathrm{GT}}(0^+_{i} \to 1_1^+)$ and $M_{\mathrm{GT}}(1_1^+ \to 0^+_{f})$ need to be fitted to GT data, ideally on GT transitions between the intermediate and the initial/final nucleus of the double-weak decay ($M_\mathrm{GT} = \sqrt{3} \, C_{n}$ with $n=\{\mathrm{EC},\beta\}$). Since this data is not available for the two decays we aim to predict, we follow a similar strategy as in Ref.~\cite{CoelloPerez:2018ghg}, determining hypothetical values for comparative half-lives: $\log ft_{\mathrm{EC}/\beta} = 5.015(120)/5.64(58)$ for $^{126}$I and $\log ft_{\mathrm{EC}/\beta} = 5.3565(5275)/4.967(134)$ for $^{134}$Cs. The LECs are then fitted to these hypothetical $\log ft$ values, $\log ft=\log(\kappa (2J_i+1)/(g_A M_\mathrm{GT})^2)$, with $\kappa = 6147\,$s the $\beta$ decay constant and $J_i$ the initial spin. 

The hypothetical $\log ft$ values are determined in the following way and encompass three sources of uncertainty:
\begin{enumerate}
    \item Experimental uncertainty in $\log ft$ values: \\
        The hypothetical range of GT EC comparative half-lives reproduces the range of $\log ft_\mathrm{EC}$ values found in the isotopic chain of the intermediate nucleus, i.e., iodine (cesium) for $^{126}$Xe ($^{134}$Xe).
    \item Ratio $r$ ($\log ft _{\beta^-} = r\,\log ft _\mathrm{EC}$): \\
        The hypothetical $\log ft_\beta$ range is deduced by applying a ratio $r$ to the $\log ft _\mathrm{EC}$ range from the previous step. This ratio is taken from two neighboring nuclei, $^{128}$I and $^{130}$Cs, which have a $1^+_\mathrm{gs}$ and decay via both EC and $\beta^-$. 
    \item Theoretical uncertainty of $\log ft$ values: \\
        Since the comparative half-life is not a purely experimental quantity, there is a theoretical uncertainty inherent to it. To estimate this, we perform the above explained fitting strategy with comparative half-lives from the \texttt{LogFT}~\cite{logft_code_LogFT} and the \texttt{BetaShape} code~\cite{logft_code_BetaShape, Mougeot_2019, Mougeot_2023}.
\end{enumerate}

\begin{figure}[t!]
    \centering
    \includegraphics[width=\linewidth]{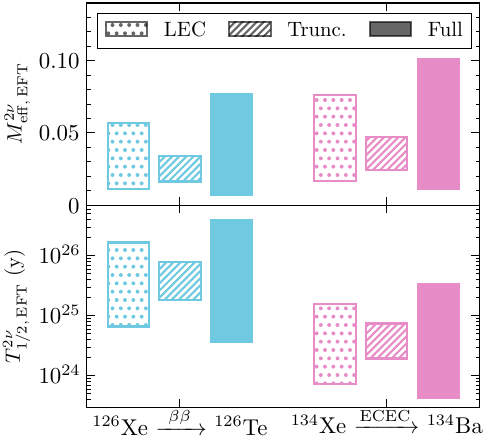}
    \caption{EFT prediction of $M^{2\nu}_\mathrm{eff}$ and $T^{2\nu}_{1/2}$ for $2\nu$ECEC of $^{126}$Xe (blue) and $2\nu\beta\beta$ decay of $^{134}$Xe (pink). Dotted and slashed bars indicate the LEC uncertainty and the EFT truncation (Trunc.) uncertainty, respectively. Solid bars correspond to the full prediction with combined uncertainties.}
    \label{fig:EFT}
\end{figure}

\begin{table*}[t!]
    \centering
    \caption{Predicted effective NMEs $M^{2\nu}_\mathrm{eff}$ (column four) and half-lives $T^{2\nu}_{1/2}$ (column five) for the ground-state-to-ground-state double-weak decays of $^{126}$Xe and $^{134}$Xe (column one). The many-body method is given in column three together with the PSF used (column two). PSF values are given in Table~\ref{tab:quantities_we_all_use}.
    }
    \begin{tabular}{ccccl}
    \toprule
    Nucleus & \multirow{2}{*}{PSF} & \multirow{2}{*}{Method} & \multirow{2}{*}{$M^{2\nu}_\mathrm{eff}$} & \multicolumn{1}{c}{ \multirow{2}{*}{$T^{2\nu}_{1/2}$($10^{24}\,$y)}}\\
    Decay mode& & & \\
    \hline
    \multirow{5}{*}{$^{126}$Xe} &\multirow{3}{*}{HSD}& pnQRPA&$0.034-0.092$&$\hphantom02.57\hphantom0-\hphantom{0}18.5$\\ 
    \multirow{5}{*}{$2\nu\mathrm{ECEC}$}&& NSM & $0.024-0.044$ & $11.2\hphantom{00}-\hphantom{0}38.1$ \\
    &&IBM-2 &$0.037-0.209$&$\hphantom00.50\hphantom0-\hphantom{0}16.2$\\[2mm]
    &\multirow{2}{*}{SSD} &IBM-2 &$0.263-1.500$&$\hphantom00.01\hphantom0-\hphantom{00}0.31$\\
    &&EFT & $0.007 - 0.077$& $\hphantom03.62\hphantom0 - 398$\\
    \hline
    \multirow{5}{*}{$^{134}$Xe} &\multirow{3}{*}{HSD}&pnQRPA &$0.034-0.060$&$\hphantom01.22\hphantom{0}-\hphantom{00}3.80$\\ 
    \multirow{5}{*}{ $2\nu\beta\beta$} & &NSM& $0.023-0.047$ & $\hphantom02.04\hphantom{0}-\hphantom{00}8.04$ \\
    &&IBM-2 &$0.040-0.234$&$\hphantom00.081-\hphantom{00}2.74$\\[2mm]
    &\multirow{2}{*}{SSD}&IBM-2 &$0.286-1.666$&$\hphantom00.002-\hphantom{00}0.053$\\
    &&EFT &$0.011 - 0.101$ &$\hphantom00.424 - \hphantom034.3$ \\
    \bottomrule
    \end{tabular} 
    \label{tab:prediction_summary}
\end{table*}

Figure~\ref{fig:EFT} shows that the largest source of uncertainty is the LEC one, which is dominated by the large uncertainty in the ratio $r$ to estimate a hypothetical $\log ft_{\beta^-}$ value. It could be significantly reduced by measuring the GT strength $B(\mathrm{GT})$ between the $1^+_1$ state of the intermediate nucleus (here $^{126}$I and $^{134}$Cs) and the corresponding initial and final nuclei of the decay (here $^{126}$Xe/$^{134}$Xe and $^{126}$Te/$^{134}$Ba). For more details, see \ref{app:EFT}.

\section{Results and discussion}
\label{sec:res_and_disc}

Our results for the effective NMEs and half-life predictions for the $^{126}$Xe $2\nu$ECEC to $^{126}$Te and the $2\nu\beta\beta$ decay of $^{134}$Xe to $^{134}$Ba using the pnQRPA, NSM, IBM-2, and EFT are presented in Table~\ref{tab:prediction_summary}.
Within each method considered in this work, the order of magnitude of the effective NMEs is common for $^{126}$Xe and $^{134}$Xe. Due to the different PSFs, the half-life predictions for $^{126}$Xe are systematically almost an order of magnitude longer than those for $^{134}$Xe.

Figure~\ref{fig:T12_allmtheods} also presents the half-life predictions compared to experimental lower limits. All half-life predictions agree within uncertainties, except for the IBM-2 SSD case. Half-lives predicted with the EFT reach up to the longest half-lives for both processes, while the IBM-2 ones are the shortest; pnQRPA and the NSM lie between them. 

The ranges of the half-lives can be very large, covering more than an order of magnitude for each of the two PSFs considered in the IBM-2. Uncertainties are also very large for the EFT calculations, which are the only ones in our work where the theoretical uncertainty can be systematically predicted within its framework. Indeed, because of the large EFT uncertainty, it fully covers the NSM (and for $^{134}$Xe, also the pnQRPA) prediction. The uncertainty of the EFT stems partially from the EFT truncation, but mainly from the LECs, see Fig.~\ref{fig:EFT}. This dominant source of uncertainty could be strongly reduced by the measurement of GT strengths. 

For the pnQRPA, the main uncertainty is driven by the ranges of the quenching factor and the associated values of the proton-neutron pairing parameter $g_{\rm pp}$ (see Fig.~\ref{fig:126134gpp} in \ref{app:pnQRPA} for illustration).
For the NSM, part of the uncertainty in $M^{2\nu}_{\rm{eff}}$ stems from the interaction uncertainty, but the dominant source is the uncertainty of the quenching factor $q$, as discussed in detail in Sec.~\ref{sec:many-body_methods}.
In IBM-2, two separate ranges are reported because the decay has been evaluated under both the HSD and SSD PSFs. While the main IBM-2 error is the range including these two possibilities, within each of them the main uncertainty also stems from the large quenching uncertainty (see Table~\ref{tab:IBM2-NMEs} in \ref{app:IBM}). 

Our results are also consistent with the values predicted by the semi-empirical formula (SEF)~\cite{Nitescu2025}: $M^{2\nu}_{\rm SEF}=0.022$ for $^{126}$Xe and $M^{2\nu}_{\rm SEF}=0.033$ for $^{134}$Xe. These correspond to half-lives $T^{2\nu}_{1/2}(^{126}{\rm Xe})=4.61\times 10^{25}$\,y and $T^{2\nu}_{1/2}(^{134}{\rm Xe})=4.03\times 10^{24}$\,y computed with the HSD PSFs listed in Table~\ref{tab:quantities_we_all_use}. For $^{126}{\rm Xe}$, our results also cover previous PHFB calculations giving $(5.7-14)\times 10^{24}$\,y~\cite{Singh:2007jh} and $(47-120)\times 10^{24}$\,y~\cite{Shukla07}.  
For $^{134}$Xe, our ranges cover previous pnQRPA results $(0.7-2.2)\times 10^{24}$\,y~\cite{Pirinen2015}, $4.57\times 10^{24}$\,y~\cite{Delion:2017bie}, and $(3.49-3.75)\times 10^{24}$\,y~\cite{Raduta:2005be}.
Our results are also consistent with previous IBM-2 calculation for $^{134}$Xe~\cite{Barea:2015kwa}, which reports only the general closure-approximation result and does not provide separate predictions for the SSD and HSD assumptions.
Finally, the differences among the various calculations arise primarily from the adopted NMEs, with minor contributions from variations in the PSFs.

\begin{figure}[t!]
    \centering
    \includegraphics[width=\linewidth]{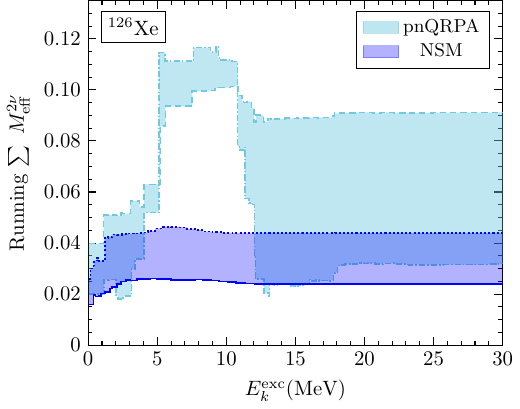}
    \caption{Running sum of the effective $2\nu {\rm ECEC}$ NME of $^{126}$Xe as a function of the excitation energy of the intermediate state, $E_k^{\rm exc}$. 
    For the pnQRPA results (in light blue), the lower (upper) limit, shown as dash-dotted (dashed) line, is obtained with the largest (smallest) value of the parameter $g_{\rm pp}^{T=0}$ (see \ref{app:pnQRPA}) with the minimum (maximum) quenching factor. The NSM results (in dark blue) are calculated with the GCN5082~\cite{Caurier:2010az} and QX~\cite{QiQX}  interactions. The lower (upper) limit, shown as a solid (dotted) line, corresponds to the GCN5082 (QX) result using the minimum (maximum) quenching factor. 
    }
    \label{fig:126running}
\end{figure}

\begin{figure}[t!]
    \centering
    \includegraphics[width=\linewidth]{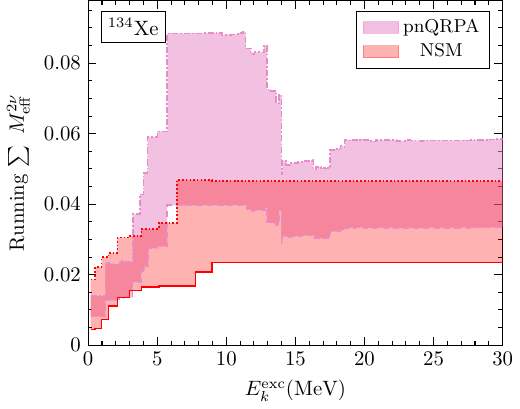}
    \caption{Same as Fig.~\ref{fig:126running} but for the effective $2\nu {\beta\beta}$ NME of $^{134}$Xe. pnQRPA results are in pink and NSM ones in red. The ranges shown for each method have the same meaning as in Fig.~\ref{fig:126running}.
    }
    \label{fig:134running}
\end{figure}

Additionally, we compare the pnQRPA and NSM running sums for $^{126}$Xe and $^{134}$Xe in Figs.~\ref{fig:126running} and~\ref{fig:134running}, respectively. The running sum reveals at which excitation energies of the intermediate nucleus there are contributions to $M^{2\nu}_{\rm eff}$. For the NSM, these contributions extend up to $\approx 5$\,MeV ($\approx10$\,MeV) in the case of $^{126}$Xe ($^{134}$Xe) and all contributions are constructive. In contrast, for the pnQRPA, contributions extend up to $\approx15$\,MeV or higher energies, and especially states between $10-15$\,MeV cancel dramatically the value of the effective NME. However, since below these energies the pnQRPA NME running sum reaches a much larger value than the NSM one, the final NMEs are similar for both many-body methods. This behavior is analogous to the previously observed running sums of the $2\nu\beta\beta$ decay of $^{136}$Xe for the same two methods~\cite{KamLAND-Zen2019}.

Figure~\ref{fig:T12_allmtheods} shows that current experimental limits for the half-lives are consistent with the theoretical predictions for both decays, except for calculations under the IBM-2 SSD case, which is disfavored especially for $^{134}$Xe. Regarding the rest of the predictions, we highlight a projected half-life sensitivity limit $1.7\times10^{24}$\,y for $^{134}$Xe~\cite{LUXZEPLIN21}, which is within the range of the pnQRPA, IBM-2 (HSD) and EFT predictions, and very close to the NSM one. 

\section{Summary}

We have predicted the ground-state-to-ground-state NMEs and half-lives for the $2\nu$ECEC of $^{126}{\rm Xe}$ and the $2\nu\beta\beta$ decay of $^{134}{\rm Xe}$ using four different many-body methods: the pnQRPA, NSM, IBM-2, and an EFT for heavy nuclei, including estimates of the theoretical uncertainties. Altogether, the calculated half-lives span the range of $\approx 10^{22}-10^{26}$\,y for $^{126}{\rm Xe}$ and $\approx 10^{21}-10^{25}$\,y for $^{134}{\rm Xe}$, and if we exclude the IBM-2 SSD case they reduce to $\approx 10^{24}-10^{26}$\,y for $^{126}{\rm Xe}$ and $\approx 10^{23}-10^{25}$\,y for $^{134}{\rm Xe}$. Our predictions agree within uncertainties (except for IBM-2 SSD) and all four many-body approaches overlap in half-live values around $10^{25}$\,y for $^{126}{\rm Xe}$ and $2\times10^{24}$\,y for $^{134}{\rm Xe}$. The current most stringent lower limit on the $^{126}{\rm Xe}$ half-life is $1.9\times 10^{22}$\,y~\cite{XMASScoll18}, and the present lower limit for $^{134}{\rm Xe}$ is $2.8\times 10^{22}$\,y~\cite{Yan24}. In both cases these are below most of our theoretical predictions---again, excluding IBM-2 SSD---especially for $^{126}{\rm Xe}$. Moreover, current searches that are still taking data will improve their sensitivity~\cite{Yan24} and the projected sensitivity in future searches for $^{134}{\rm Xe}$ reaches up to $1.7\times10^{24}$\,y~\cite{LUXZEPLIN21}. If this sensitivity is achieved by any current or future experiment, they will be able to probe part of the theoretical predictions in this work. This can provide valuable benchmarks for NME calculations, as well as tests of the different many-body approaches.
\begingroup
\renewcommand{\thefigure}{A.\arabic{figure}}
\begin{figure*}[p!]
    \centering
    \includegraphics[width = 0.93\linewidth]{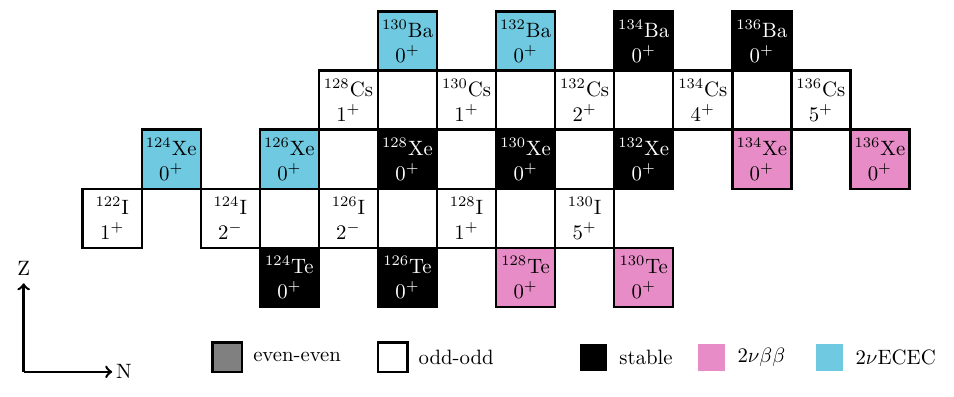}
    \caption{Part of the nuclear chart highlighting xenon and neighboring isotopes taking part in $2\nu$ double-weak decays as the initial, intermediate, or final nucleus. Even-even (odd-odd) isotopes are colored (blank) squares and odd-mass isotopes are not shown. Even-even isotopes are colored in NNDC blue (NNDC pink) to show which nuclei are energetically allowed to $2\nu$ECEC ($2\nu\beta\beta$ decay), i.e., $Q_{\mathrm{EC}/\beta}<0<Q_{\mathrm{2EC}/2\beta}$. Black isotopes are stable.}
    \label{fig:app_nuclchart}
\end{figure*}
\endgroup

\section*{Acknowledgements}

We thank O.~Ni\c tescu for helpful discussions regarding the semi-empirical formula. This work was supported in part by the European Research Council (ERC) under the European Union’s Horizon 2020 research and innovation programme (Grant Agreement No.~101020842) and by the LOEWE Top Professorship LOEWE/4a/519/05.00.002(0014)98 by the State of Hesse; by the Romanian Ministry of Research, Innovation and Digitalization under the project PNRR-I8/C9-CF264, Contract No.~60100/23.5.2023 (the NEPTUN project); the Office of Nuclear Physics, U.S. Department of Energy, under Grant No.~DE-FG02-97ER41019; and by the Spanish Ministry of Science and Innovation MCIN/AEI/10.13039/501100011033 from the following grants: PID2023-147112NB-C22, CNS2022-135716 funded by the “European Union NextGenerationEU/PRTR”, and CEX2024-001451-M to the “Unit of Excellence María de Maeztu 2025-2031” award to the Institute of Cosmos Sciences.

\appendix

\section{Input for double-weak decays}
\label{app:all}

\begin{table}[t!]
    \centering
    \caption{Summary of quantities used by all many-body methods. Column one and two specify the nucleus and the related quantity. Column three and four give the value and the corresponding reference. $G^{2\nu}$ is the PSF and its subscript corresponds to the SSD or HSD case. The binding energy per nucleon (excitation energy of first $1^+_1$ state) is denoted as $B/A$ ($E_{1^+_1}$). }
    \begin{tabular}{l@{\hspace{8pt}}c@{\hspace{8pt}}l@{\hspace{8pt}}c}
    \toprule
        Nucl. & Quant. & \multicolumn{1}{l}{value in units} & Ref.\\\hline
        \multirow{3}{*}{$^{126}$Xe}&$G^{2\nu}_\mathrm{SSD}$ & \hphantom{111}$4.69\times10^{-23}$\,$\mathrm{y}^{-1}$ &this work\\
        &$G^{2\nu}_\mathrm{HSD}$ &\hphantom{111}$4.61\times10^{-23}$\,$\mathrm{y}^{-1}$ &\cite{Kotila:2013gea}\\
        & $B/A$ & $8443.538$\,keV & \cite{Wang_2021} \\[1mm]
        \multirow{2}{*}{$^{126}$I} & $E_{1^+_1}$ & \hphantom{11}54.43(4)\,keV & \cite{ensdf}\\
        & $B/A$ & $8439.938$\,keV & \cite{Wang_2021} \\[1mm]
        $^{126}$Te & $B/A$ & $8463.240$\,keV & \cite{Wang_2021}\\
        \midrule
        \multirow{3}{*}{$^{134}$Xe}&$G^{2\nu}_\mathrm{SSD}$ & \hphantom{111}$2.31\times10^{-22}$\,$\mathrm{y}^{-1}$ &this work\\
        &$G^{2\nu}_\mathrm{HSD}$ & \hphantom{111}$2.26\times10^{-22}$\,$\mathrm{y}^{-1}$ &\cite{Kotila2012}\\
        &$B/A$& $8413.699$\,keV & \cite{Wang_2021}\\[1mm]
        \multirow{2}{*}{$^{134}$Cs} & $E_{1^+_1}$& \hphantom{1}176.6400(26)\,keV  & \cite{ensdf}\\
        & $B/A$ & $8398.647$\,keV & \cite{Wang_2021}\\[1mm]
        $^{134}$Ba & $B/A$ & $8408.173$\,keV & \cite{Wang_2021}\\
        \bottomrule
    \end{tabular}
    \label{tab:quantities_we_all_use}
\end{table}

Figure~\ref{fig:app_nuclchart} shows the part of the nuclear chart that contains the even-even xenon isotopes involved in $2\nu\beta\beta$ or $2\nu$ECEC decays. Following the color scheme of NNDC~\cite{nudat3}, stable isotopes are filled with black, and isotopes that decay via $2\nu\beta\beta$ decay ($2\nu$ECEC) are filled with pink (light blue). Neighboring odd-odd isotopes, that we use in some of our methods, are left blank.

Table~\ref{tab:quantities_we_all_use} summarizes the experimental values for the binding energies and excitation energies of the lowest $1^+_1$ states in the intermediate nuclei, as well as the PSFs $G^{2\nu}$ we use. The PSFs are based on the formalism of Refs.~\cite{Kotila2012, Kotila:2013gea}, employing exact Dirac electron wave functions with finite nuclear size and electron screening effects. The PSFs are evaluated by integrating over the lepton phase space: a four-body continuum for \(2\nu\beta\beta\), and bound-state electron wave functions at the nucleus for \(2\nu\)ECEC. To enable the factorization of the PSF and NME, the excitation energies of the intermediate $1^+$ states in the energy denominator are approximated by an average excitation energy in the HSD case, or by the excitation energy of the lowest $1^+_1$ state in the SSD one. As a result, the SSD approach yields a different energy dependence and different numerical values than the HSD approximation.

\section{pnQRPA}
\label{app:pnQRPA}

\begin{figure*}[p!]
    \centering
    \includegraphics[width=0.49\linewidth]{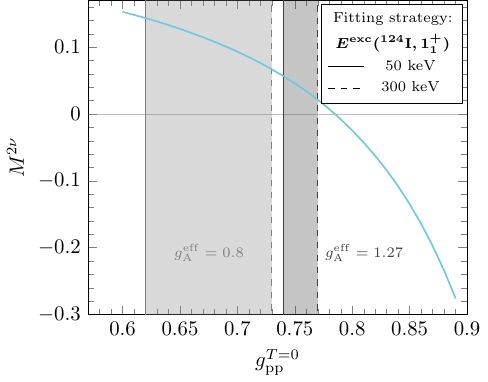}
    \includegraphics[width=0.49\linewidth]{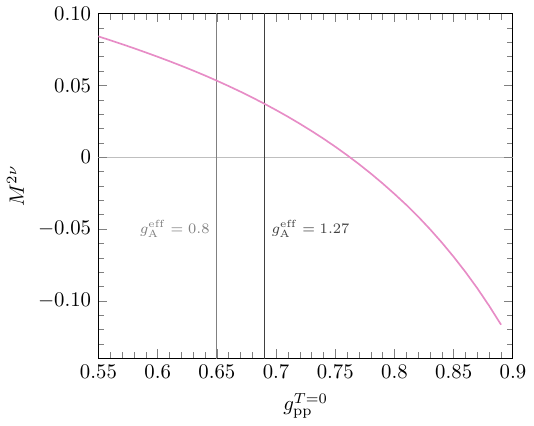}
    \caption{Left: $2\nu {\rm ECEC}$ matrix element of $^{126}$Xe as a function of the particle-particle pairing parameter $g_{\rm pp}^{T=0}$. The vertical lines denote the values fitted to the measured half-life of $2\nu{\rm ECEC}$ of $^{124}$Xe assuming the effective axial-vector coupling $g_{\rm A}^{\rm eff}=qg_{\rm A}=0.8$ and $g_{\rm A}^{\rm eff}=1.27$ for the range of $E^{\rm exc}(^{124}{\rm I},1^+_1)$. Right: $2\nu\beta\beta$ decay NME of $^{134}$Xe as a function of the particle-particle pairing parameter $g_{\rm pp}^{T=0}$. Vertical lines denote the values fitted to the measured half-life of $2\nu\beta\beta$ decay of $^{136}$Xe assuming the effective axial-vector constants $g_{\rm A}^{\rm eff}=qg_{\rm A}=0.8$ and $g_{\rm A}^{\rm eff}=1.27$.}
    \label{fig:126134gpp}
\end{figure*}

\begin{table*}[p!]
    \caption{pnQRPA predictions for the NMEs and half-lives for $2\nu$ECEC of $^{126}$Xe and $2\nu\beta\beta$ decay of $^{134}$Xe. For $^{126}$Xe, the range in $M^{2\nu}$ results from the unknown energy of the lowest $1^+_1$ state in $^{124}$I used in the fitting of the $g_{\rm pp}^{T=0}$ parameter.}
    \label{tab:pnQRPA-nmes}
    \centering
    \begin{tabular}{cccccccc}
    \toprule
    Nucleus & \multirow{2}{*}{PSF} & \multirow{2}{*}{$M^{2\nu}$} &\multirow{2}{*}{$g_{\rm A}^{\rm eff}=qg_{\rm A}$}& \multirow{2}{*}{$M^{2\nu}_{\rm eff}$ }&\multirow{2}{*}{$T^{2\nu}_{1/2} ({\rm y})$}\\
    Decay mode\\
    \midrule
    $^{126}$Xe& \multirow{2}{*}{HSD} &$0.067-0.144$  &0.8\hphantom0 &$0.043-0.092$ &$(0.257-1.18)\times 10^{25}$ \\
    $2\nu$ECEC &  &$0.021-0.057$ &1.27&$0.034-0.092$ &$(0.259-1.85)\times 10^{25}$ \\
    \midrule
    $^{134}$Xe& \multirow{2}{*}{HSD}   &0.053&0.8\hphantom0 &0.034 &$3.81\times 10^{24}$\\
    $2\nu\beta\beta$& &0.037 &1.27 &0.060 &$1.22\times 10^{24}$\\
    \bottomrule
    \end{tabular}
\end{table*}

Figure~\ref{fig:126134gpp} shows the NMEs for $^{126}$Xe and $^{134}$Xe as functions of the isoscalar particle-particle parameter $g_{\rm pp}^{T=0}$. The isovector part, $g_{\rm pp}^{T=1}$, is adjusted so that the Fermi part of the NME vanishes, following the partial isospin restoration scheme~\cite{Simkovic2013}. The vertical gray lines show the values of $g_{\rm pp}^{T=0}$ fitted to the measured $2\nu \rm ECEC$ in $^{124}$Xe (left panel of Fig.~\ref{fig:126134gpp}) and $2\nu\beta\beta$ decay in $^{136}$Xe (right panel of Fig.~\ref{fig:126134gpp}), assuming effective axial-vector couplings $g_{\rm A}^{\rm eff}=0.8$ and 1.27. For $^{126}$Xe, the shaded areas correspond to the variation of the energy of the lowest intermediate $1^+$ state within the range of $E^{\rm exc}(^{124}{\rm I},1^+_1)=50-300$ keV, the solid (dashed) line denoting $E^{\rm exc}(^{124}{\rm I},1^+_1)=50$\,keV ($E^{\rm exc}(^{124}{\rm I},1^+_1)=300$\,keV).

The computed NMEs and half-lives with different choices of effective axial-vector coupling are listed in Table~\ref{tab:pnQRPA-nmes}.

\section{NSM}
\label{app:NSM}

\begin{table*}[t!]
    \centering
    \caption{NSM predictions for NMEs of $2\nu$ECEC of $^{126}$Xe using the GCN5082 and QX effective interactions as a function of increasing jump truncations used in the shell-model diagonalization.}
    \begin{tabular}{cccccc}
    \toprule
    Nucleus & \multirow{2}{*}{$H_{\rm eff}$}  & \multicolumn{3}{c}{$M^{2\nu}_{\rm eff}$}\\
    Decay mode && jump\,=\,0 & jump\,=\,2 & jump\,=\,3\\
    \midrule
    $^{126}$Xe &GCN5082 & $0.061-0.113$ & $0.049-0.091$ & $0.024-0.044$ \\
    $2\nu$ECEC &QX & $0.045-0.067$ & $0.040-0.060$ & $0.025-0.038$ \\
    \bottomrule
    \end{tabular}
    \label{tab:nsm_xe126_trunc}
\end{table*}

\begin{table*}[t!]
    \centering
    \caption{NSM predictions for the NMEs and half-lives for $2\nu$ECEC of $^{126}$Xe and $2\nu\beta\beta$ decay of $^{134}$Xe using the GCN5082 and QX effective interactions. Quenching factors are in the ranges $q=0.42-0.57$ and $q=0.67-0.82$ for GCN5082 and QX, respectively. For $^{126}$Xe we allow up to three nucleons excited from $1d_{5/2}$, $0g_{7/2}$ to $2s_{1/2}$, $1d_{3/2}$, $0h_{11/2}$ orbitals (jump\,=\,3).}
    \begin{tabular}{ccccc}
    \toprule
    Nucleus & \multirow{2}{*}{PSF}
    & \multirow{2}{*}{$H_{\rm eff}$} & \multirow{2}{*}{$M^{2\nu}_{\rm eff}$} & \multirow{2}{*}{$T^{2\nu}_{1/2} (\mathrm{y})$ }\\
    Decay mode &\\
    \midrule
    \multirow{1}{*}{$^{126}$Xe} & \multirow{2}{*}{HSD} & GCN5082 & $0.024-0.044$ & $(1.12-3.82)\times 10^{25}$ \\
    \multirow{1}{*}{$2\nu$ECEC} & & QX & $0.025-0.038$ & $(1.52-3.43)\times 10^{25}$ \\
    \midrule
    \multirow{1}{*}{$^{134}$Xe} & \multirow{2}{*}{HSD} & GCN5082 & $0.025-0.047$ & $(2.04-6.93)\times10^{24}$ \\
    \multirow{1}{*}{$2\nu\beta\beta$}& & QX & $0.023-0.035$ & $(3.58-8.04)\times 10^{24}$ \\
    \bottomrule
    \end{tabular}
    \label{tab:nsm_nmes}
\end{table*}

\begin{figure}[t!]
    \centering
    \includegraphics[width=\linewidth]{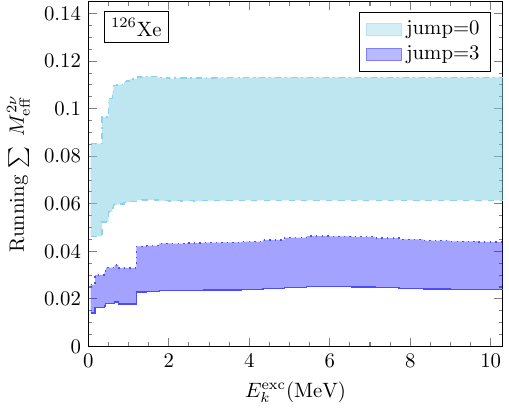}
    \caption{Running sum of the effective $2\nu {\rm ECEC}$ NME of $^{126}$Xe as a function of the excitation energy of the intermediate state, $E_k^{\rm exc}$, computed for the GCN5082 interaction with two different truncations. The ranges of the bands correspond to the values given in Table~\ref{tab:nsm_nmes} after taking into account the quenching factor range for each truncated configuration space.}
    \label{fig:126xe_nsm_trunc}
\end{figure}

Due to the large dimensions of the many-body spaces to be diagonalized for the relevant $0^+$ and $1^+$ states, the calculation of $2\nu$ECEC NMEs of $^{126}$Xe demands a truncation of the configuration space. We follow Ref.~\cite{Nitescu24} and characterize the truncation by a parameter ``jump'', which represents the number of nucleons allowed to be excited from the lower-energy orbitals $g_{7/2}d_{5/2}$ to the higher-energy orbitals $2s_{1/2}$, $1d_{3/2}$, and $0h_{11/2}$. Table~\ref{tab:nsm_xe126_trunc} shows the NSM predictions for the $2\nu$ECEC NMEs of $^{126}$Xe as a function of the configuration space truncation. As we increase the configuration space to allow additional correlations between nucleons, the NME is reduced, similar to the behaviour of the $2\nu$ECEC NME of $^{124}$Xe. Note that we use the $\text{jump}=3$ for our final NSM NME results, but do not include an uncertainty from this truncation.

Figure~\ref{fig:126xe_nsm_trunc} shows the running sum of the effective $2\nu {\rm ECEC}$ NME of $^{126}$Xe computed for the GCN5082 interaction with two different truncations, $\text{jump}=0$ and $\text{jump}=3$. This figure indicates that the dominant change when increasing to $\text{jump}=3$ comes from the contribution of the lowest-lying intermediate $1^+_1$ state, while higher-state contributions exhibit a similar behavior of the running sum in both truncations.

Finally, Table~\ref{tab:nsm_nmes} summarizes the computed NMEs and half-lives for the $2\nu\beta\beta$-decay of $^{134}$Xe and the $2\nu$ECEC of $^{126}$Xe in the NSM.

\section{IBM-2}
\label{app:IBM}

\begin{table*}[t!]
    \centering
    \caption{Parameters for the IBM-2 Hamiltonian.}
    \begin{tabular}{lcccccccccc}
    \toprule
    Nucleus & $\varepsilon_d$ & $\kappa$ & $\chi_\nu$ & $\chi_\pi$ & $\xi_1$ & $\xi_2$ & $\xi_3$ & $c^{(0)}_\nu$ & $c^{(2)}_\nu$ & Ref. \\
    \midrule
       $^{126}$Xe & 0.643 & $-0.228$ & 0.30 & $-0.30$ & 0.30 & 0.12 & 0.30 & 0.02 & $-0.10$ & \cite{GADE2000268} \\
       $^{126}$Te & 0.860 & $-0.130$ & 0.20 & $-1.60$ & $-0.18$ & 0.24 & $-0.18$ & 0.30 & 0.22 & this work \\
       $^{134}$Xe & 1.030 & $-0.230$ & 1.00 & $-0.80$ & 0.00 & 0.00 & 0.00 & 0.30 & 0.10 & \cite{GADE2000268} \\
       $^{134}$Ba & 0.880 & $-0.200$ & 0.79 & $-0.90$ & 0.00 & 0.00 & 0.00 & 0.30 & 0.10 & \cite{PUDDU1980109} \\       
    \bottomrule
    \end{tabular}
    \label{tab:IBM2-parameters}
\end{table*}

\begin{table*}[t!]
    \centering
    \caption{IBM-2 predictions of NMEs and half-lives calculated for $2\nu$ECEC of $^{126}$Xe and $2\nu\beta\beta$ decay of $^{134}$Xe. The values are computed with three different $g_{\rm A}^{\rm eff}$ values, obtained from a fit to experimental $2\nu\beta\beta$ decay data in previous IBM-2 studies.}
    \begin{tabular}{cccccc}
    \toprule
    Nucleus &\multirow{2}{*}{PSF} & \multirow{2}{*}{$M^{2\nu}$}& \multirow{2}{*}{$g_{\rm A}^{\rm eff}=qg_{\rm A}$} & \multirow{2}{*}{$M^{2\nu}_{\rm eff}$} &\multirow{2}{*}{$T^{2\nu}_{1/2}\,({\rm y})$}\\
    Decay mode & & & & & \\
    \midrule
    \multirow{5}{*}{$^{126}$Xe} &\multirow{3}{*}{HSD}  & \multirow{3}{*}{0.130} &1.269 & 0.209 & $4.97\times10^{23}$ \\
    \multirow{5}{*}{$2\nu$ECEC}& & &1.00\hphantom0 & 0.130 & $1.28\times 10^{24}$ \\
    & & &$1.269A^{-0.18}$& 0.037 & $1.61\times 10^{25}$\\[2mm]
    & \multirow{3}{*}{SSD} & \multirow{3}{*}{0.931} &$1.269$& 1.500 & $9.48\times 10^{21}$ \\
    & & &1.00\hphantom0 & 0.931 & $2.45\times 10^{22}$ \\
    & & &$1.269A^{-0.18}$& 0.263 & $3.09\times 10^{23}$ \\
    \midrule
    \multirow{5}{*}{$^{134}$Xe} & \multirow{3}{*}{HSD} & \multirow{3}{*}{0.145}& 1.269 & 0.234 & $8.07\times10^{22}$ \\
    \multirow{5}{*}{$2\nu\beta\beta$}& & & 1.00\hphantom0& 0.145 & $2.09\times10^{23}$\\
    & & & $1.269A^{-0.18}$ & 0.040 & $2.74\times10^{24}$\\[2mm]
    & \multirow{3}{*}{SSD} & \multirow{3}{*}{1.035}& 1.269 & 1.666 & $1.56\times10^{21}$\\
    & & & 1.00\hphantom0 & 1.035 & $4.04\times10^{21}$\\
    & & & $1.269A^{-0.18}$ & 0.286 & $5.30\times10^{22}$\\
    \bottomrule
    \end{tabular}
    \label{tab:IBM2-NMEs}
\end{table*}

The parameters for the IBM-2 Hamiltonian are presented in Table~\ref{tab:IBM2-parameters}, where the parameters for $^{126}$Te are determined in this work. The computed IBM-2 NMEs and half-lives with different choices of effective axial-vector coupling and HSD/SSD PSFs are listed in Table~\ref{tab:IBM2-NMEs}.

\section{EFT for heavy nuclei}
\label{app:EFT}

Section~\ref{subsec:EFT} presents the hypothetical $\log ft$ values to which the EFT LECs are fitted. This appendix details how these values are obtained. 

For hypothetical EC $\log ft$ values, we use the range of values found in the isotopic chain of the intermediate nucleus in the double-weak decays. 
As can be seen in Fig.~\ref{fig:app_nuclchart}, the intermediate nuclei are $^{126}$I ($2\nu$ECEC of $^{126}$Xe) and $^{134}$Cs ($2\nu\beta\beta$ of $^{134}$Xe). Table~\ref{tab:EFT_iso_chain_logfts} gives the iodine and cesium isotopes with a $1^+_{\text{gs}}$ ground-state (gs) and an available $\log ft$ value~\cite{nudat3}.

For iodine, as in Ref.~\cite{CoelloPerez:2018ghg}, we only use the values of $^{122,128}$I. We discard $^{116}$I, because there is no uncertainty reported and it has 10 neutrons less than the isotope of interest $^{126}$I. For cesium, we use $^{128,130}$Cs with the following reasoning. 
In the EFT, nucleons and phonons are coupled to a spherical even-even core. Thus, a $2\nu$ double-weak process with a deformed initial and deformed final nuclei is not well described by the EFT. A spherical nucleus exhibits an excitation-energy ratio of the first $4^+_1$ and $2^+_1$ states of $E_{4^+_1}/E_{2^+_1} = 2$. We thus require, like in previous EFT works, that $1.5\leqslant E_{4^+_1}/E_{2^+_1}\leqslant 2.5$. 

\begin{figure}[t!]
    \centering
    \includegraphics[width=\linewidth]{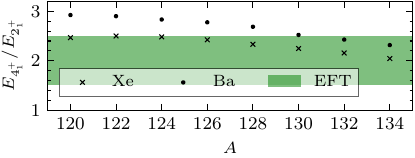}
    \caption{Excitation-energy ratio of the first $4^+_1$ and $2^+_1$ states for xenon (black crosses) and barium (black dots) isotopes for mass number $120 \leqslant A \leqslant 134$, which is compared to the EFT range (green bar). Data taken from Ref.~\cite{ensdf}.}
    \label{fig:app_EFT_rE4p2p}
\end{figure}

While xenon isotopes meet this condition, in barium the upper limit is already exceeded at $^{130}$Ba ($E_{4^+}/E_{2^+} = 2.52$), and the ratio increases with decreasing neutron number up to $^{120}$Ba ($E_{4^+}/E_{2^+} = 2.92$) (see Fig~\ref{fig:app_EFT_rE4p2p}). So we discard $^{122,124}$Cs based on the deformation of the adjacent even-even nuclei. Additionally, $^{122,124}$Cs have ten or more neutrons less than the isotope of interest $^{134}$Cs. 
The $\beta^-$ $\log ft$ values are obtained by multiplying the hypothetical EC $\log ft$ range with a factor based on the empirical ratio of $\beta^-$ and EC $\log ft$ values observed in $^{128}$I and $^{130}$Cs. 

\begin{table}[t!]
    \centering
    \caption{Reported $\log ft$-values for ground-state to ground-state decays of $1^+_\mathrm{gs}$ iodine and cesium isotopes. Mass number and element are given in columns one and two. The decay mode (column three) and the corresponding reported $\log ft$ values can be found in column four (\texttt{LogFT}~\cite{logft_code_LogFT}) and five (\texttt{BetaShape}~\cite{logft_code_BetaShape,Mougeot_2019,Mougeot_2023}, only for the isotopes used for the LEC fits). }
    \begin{tabular}{cccll}\toprule
        \multicolumn{2}{c}{Nucleus} & Decay & $\log ft_\texttt{LogFT} $ & $\log ft_\texttt{BetaShape}$ \\ \hline
        116 & \multirow{4}{*}{\hspace{-3mm}I} & $\beta^+$ &4.5 \\
        122 & &EC& 4.95(4)& 4.964(33)\\
        \multirow{2}{*}{128} & &EC& 5.05(5) & 5.08(5)\\
        &&$\beta^-$&6.063(6) & 6.095(5)\\
        \midrule
        122 &\multirow{6}{*}{\hspace{-3mm}Cs}&EC& 5.39(5)\\
        124 & &EC& 5.10(7)\\
        126 & &EC& 5.053(19)\\
        128 & &EC& 4.843(10) & 4.868(10)\\
        \multirow{2}{*}{130} & &EC& 5.073(6) & 5.096(5)\\
        & &$\beta^-$& 5.36(6) & 5.15(9)\\
        \bottomrule
    \end{tabular}
    \label{tab:EFT_iso_chain_logfts}
\end{table}

Therefore, for $^{126}$I the EC range is taken from $^{122}$I and $^{128}$I, but for $^{134}$Cs only comparative half-lives of lighter nuclei are available. We additionally check the validity of our hypothetical $\log ft_{\beta^-}$ range for $^{134}$Cs by fitting the LEC to the GT strength from a charge-exchange reaction involving a heavier nucleus, $^{136}\mathrm{Xe}(^3\mathrm{He},t)^{136}\mathrm{Cs}$~\cite{Puppe2011}. The result leads to $M^{2\nu}_{\mathrm{EFT,\,} B(\mathrm{GT})} = 0.007-0.057$, consistent with the prediction when fitting only to the comparative half-lives of lighter isotopes, see Table~\ref{tab:prediction_summary}.

Since $\log ft$ is a product of the observable half-life $t$ and the calculated PSF $f$, we estimate the uncertainty induced by the latter. We consider $\log ft$ values calculated with \texttt{BetaShape}~\cite{logft_code_BetaShape,Mougeot_2019,Mougeot_2023} in addition to the $\log ft$ values found in Ref.~\cite{nudat3} evaluated with \texttt{LogFT}~\cite{logft_code_LogFT}. We take the range given by these two codes using different approximations in the evaluation of the PSFs.
This uncertainty varies with the specific nucleus, from $<1\%$ for $^{122}$I to $\approx 5\%$ for $\beta^-$ of $^{130}$Cs, see Table~\ref{tab:EFT_iso_chain_logfts}.

\begin{figure}[t!]
    \centering
    \includegraphics[width=\linewidth]{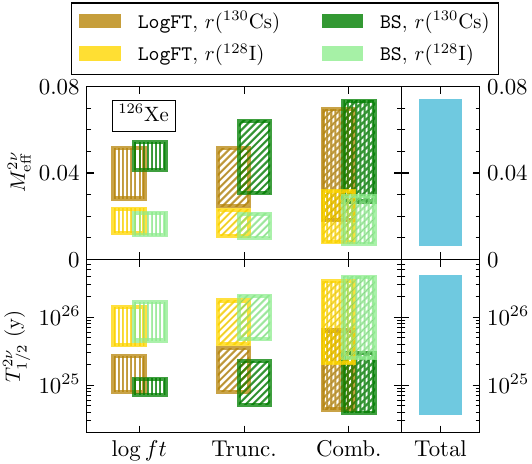}
    \caption{$2\nu$ECEC NME $M^{2\nu}_\mathrm{eff}$ (top panels) and half-life of $^{126}$Xe (bottom panels) for different fitting strategies. The left panels resolve the details and the right panels display the total prediction containing all quantified uncertainties. The different hatching styles correspond to different uncertainties included in the calculation, specified in the x-axis labels ($\log ft$, EFT truncation, and combined uncertainty). The other colors specify the fitting strategy. Lighter (darker) colors show the results when the ratio $r$ is taken from $^{128}$I ($^{130}$Cs). Golden (green) colors show the predictions when the LECs are fitted to $\log ft$ values from the \texttt{LogFT} (\texttt{BetaShape} (\texttt{BS})) code.}
    \label{fig:app_EFT_details}
\end{figure}

Figure~\ref{fig:app_EFT_details} shows the predictions using the two different ratios ($^{128}$I and $^{130}$Cs) from the two codes (\texttt{LogFT} and \texttt{BetaShape}) in the left panels and the total prediction in the right panels. The vertically (diagonally) hatched bars show the uncertainty of the hypothetical $\log ft $ ranges (the EFT truncation). The vertically and diagonally hatched bars show the full result for each of the fitting strategies. 

The predictions between \texttt{LogFT} and \texttt{BetaShape} agree within uncertainties. 
For $^{128}$I, \texttt{BetaShape} gives slightly larger values ($+0.03$) for both EC and $\beta^-$ $\log ft$-values compared to \texttt{LogFT}. Therefore, the extracted EC range is only shifted to slightly larger values and the ratio $r$ is very similar. In contrast, for $^{130}$Cs $\log ft_{\mathrm{EC},\texttt{BetaShape}}$ increases ($+0.02$) and $\log ft_{\beta^-,\texttt{BetaShape}}$ decreases ($-0.1$) compared to the corresponding values from \texttt{LogFT} (see Table~\ref{tab:EFT_iso_chain_logfts}). This leads to a different $\beta^-$-EC ratio and uncertainty range compared to \texttt{LogFT} (see Fig.~\ref{fig:app_EFT_details}).

The dominant source of uncertainty is the ratio to obtain a hypothetical $\log ft_{\beta^-}$ value. As mentioned in Sec.~\ref{subsec:EFT}, this uncertainty could be strongly reduced by measuring the GT strength~$B(\mathrm{GT})$ between the intermediate nucleus (here $^{126}$I and $^{134}$Cs) in its first $1^+_1$ state and the corresponding initial and final nuclei of the targeted decay (here $^{126}$Xe/$^{134}$Xe and $^{126}$Te/$^{134}$Ba).
Such measurement would reduce the uncertainty, e.g., to one of the four colors in the left panels. Then the EFT uncertainty could be comparable or even dominant compared to the $\log ft$/experimental ones (e.g., dark golden or dark green). 

\bibliography{biblio}

\begin{thebibliography}{10}
\expandafter\ifx\csname url\endcsname\relax
  \def\url#1{\texttt{#1}}\fi
\expandafter\ifx\csname urlprefix\endcsname\relax\def\urlprefix{URL }\fi
\expandafter\ifx\csname href\endcsname\relax
  \def\href#1#2{#2} \def\path#1{#1}\fi

\bibitem{Agostini:2022zub}
M.~Agostini, G.~Benato, J.~A. Detwiler, J.~Men{\'e}ndez, F.~Vissani, Toward the discovery of matter creation with neutrinoless {\ensuremath{\beta}}{\ensuremath{\beta}} decay, Rev. Mod. Phys. 95~(2) (2023) 025002.
\newblock \href {https://doi.org/10.1103/RevModPhys.95.025002} {\path{doi:10.1103/RevModPhys.95.025002}}.

\bibitem{Dolinski:2019nrj}
M.~J. Dolinski, A.~W.~P. Poon, W.~Rodejohann, Neutrinoless double-beta decay: Status and prospects, Annu. Rev. Nucl. Part. Sci. 69 (2019) 219--251.
\newblock \href {https://doi.org/10.1146/annurev-nucl-101918-023407} {\path{doi:10.1146/annurev-nucl-101918-023407}}.

\bibitem{Gomez-Cadenas:2023vca}
J.~J. G{\'o}mez-Cadenas, J.~Mart{\'\i}n-Albo, J.~Men{\'e}ndez, M.~Mezzetto, F.~Monrabal, M.~Sorel, The search for neutrinoless double-beta decay, Riv. Nuovo Cim. 46~(10) (2023) 619--692.
\newblock \href {https://doi.org/10.1007/s40766-023-00049-2} {\path{doi:10.1007/s40766-023-00049-2}}.

\bibitem{Saakyan:2013yna}
R.~Saakyan, Two-neutrino double-beta decay, Annu. Rev. Nucl. Part. Sci. 63 (2013) 503--529.
\newblock \href {https://doi.org/10.1146/annurev-nucl-102711-094904} {\path{doi:10.1146/annurev-nucl-102711-094904}}.

\bibitem{Barabash2020}
A.~Barabash, Precise half-life values for two-neutrino double-$\beta$ decay: 2020 review, Universe 6 (2020) 159.
\newblock \href {https://doi.org/10.3390/universe6100159} {\path{doi:10.3390/universe6100159}}.

\bibitem{Engel2017}
J.~Engel, J.~Men{\'e}ndez, Status and future of nuclear matrix elements for neutrinoless double-beta decay: a review, Rep. Prog. Phys. 80 (2017) 046301.
\newblock \href {https://doi.org/10.1088/1361-6633/aa5bc5} {\path{doi:10.1088/1361-6633/aa5bc5}}.

\bibitem{Jokiniemi:2022ayc}
L.~Jokiniemi, B.~Romeo, P.~Soriano, J.~Men{\'e}ndez, Neutrinoless {\ensuremath{\beta}}{\ensuremath{\beta}}-decay nuclear matrix elements from two-neutrino {\ensuremath{\beta}}{\ensuremath{\beta}}-decay data, Phys. Rev. C 107~(4) (2023) 044305.
\newblock \href {https://doi.org/10.1103/PhysRevC.107.044305} {\path{doi:10.1103/PhysRevC.107.044305}}.

\bibitem{Horoi:2022ley}
M.~Horoi, A.~Neacsu, S.~Stoica, Statistical analysis for the neutrinoless double-{\ensuremath{\beta}}-decay matrix element of \ensuremath{^{48}}{Ca}, Phys. Rev. C 106~(5) (2022) 054302.
\newblock \href {https://doi.org/10.1103/PhysRevC.106.054302} {\path{doi:10.1103/PhysRevC.106.054302}}.

\bibitem{Horoi:2023uah}
M.~Horoi, A.~Neacsu, S.~Stoica, Predicting the neutrinoless double-{\ensuremath{\beta}}-decay matrix element of \ensuremath{^{136}}{Xe} using a statistical approach, Phys. Rev. C 107~(4) (2023) 045501.
\newblock \href {https://doi.org/10.1103/PhysRevC.107.045501} {\path{doi:10.1103/PhysRevC.107.045501}}.

\bibitem{Horoi:2026fxp}
M.~Horoi, A.~Neacsu, Uncertainty quantification of the \ensuremath{^{76}}{Ge} neutrinoless double-beta decay nuclear matrix element (2026).
\newblock \href {http://arxiv.org/abs/2605.21657} {\path{arXiv:2605.21657}}.

\bibitem{Lian:2026avy}
X.~Lian, C.~R. Ding, C.~L. Bai, J.~M. Yao, Ab initio correlations between neutrinoless and two-neutrino double-beta decays in \ensuremath{^{48}}{Ca} (2026).
\newblock \href {http://arxiv.org/abs/2605.19479} {\path{arXiv:2605.19479}}.

\bibitem{XENON:2019dti}
E.~Aprile~({XENON collaboration}), et~al., Observation of two-neutrino double electron capture in $^{124}\mathrm{Xe}$ with {XENON1T}, Nature 568~(7753) (2019) 532--535.
\newblock \href {https://doi.org/10.1038/s41586-019-1124-4} {\path{doi:10.1038/s41586-019-1124-4}}.

\bibitem{XENON:2022evz}
E.~Aprile~({XENON collaboration}), et~al., Double-weak decays of $^{124}\mathrm{Xe}$ and $^{136}\mathrm{Xe}$ in the {XENON1T} and {XENONnT} experiments, Phys. Rev. C 106~(2) (2022) 024328.
\newblock \href {https://doi.org/10.1103/PhysRevC.106.024328} {\path{doi:10.1103/PhysRevC.106.024328}}.

\bibitem{LZ:2024wvs}
J.~Aalbers ({LUX-ZEPLIN}~collaboration), et~al., Two-neutrino double electron capture of \ensuremath{^{124}}{Xe} in the first {LUX-ZEPLIN} exposure, J. Phys. G 52~(1) (2025) 015103.
\newblock \href {https://doi.org/10.1088/1361-6471/ad9039} {\path{doi:10.1088/1361-6471/ad9039}}.

\bibitem{PandaX-4T:2024fls}
Z.~Bo~({PandaX}~collaboration), et~al., Measurement of two-neutrino double electron capture half-life of \ensuremath{^{124}}{Xe} with pandax-4t, JHEP 05 (2025) 119.
\newblock \href {https://doi.org/10.1007/JHEP05(2025)119} {\path{doi:10.1007/JHEP05(2025)119}}.

\bibitem{CoelloPerez:2018ghg}
E.~A. Coello~P\'erez, J.~Men\'endez, A.~Schwenk, Two-neutrino double electron capture on $^{124}\mathrm{Xe}$ based on an effective theory and the nuclear shell model, Phys. Lett. B 797 (2019) 134885.
\newblock \href {https://doi.org/10.1016/j.physletb.2019.134885} {\path{doi:10.1016/j.physletb.2019.134885}}.

\bibitem{Pirinen2015}
P.~Pirinen, J.~Suhonen, Systematic approach to $\beta$ and $2\nu\beta\beta$ decays of mass {$A=100-136$} nuclei, Phys. Rev. C 91 (2015) 054309.
\newblock \href {https://doi.org/10.1103/PhysRevC.91.054309} {\path{doi:10.1103/PhysRevC.91.054309}}.

\bibitem{Suhonen:2013rca}
J.~Suhonen, Double beta decays of \ensuremath{^{124}}{Xe} investigated in the {QRPA} framework, J. Phys. G 40 (2013) 075102.
\newblock \href {https://doi.org/10.1088/0954-3899/40/7/075102} {\path{doi:10.1088/0954-3899/40/7/075102}}.

\bibitem{KamLAND-Zen:2024eml}
S.~Abe ({KamLAND-Zen}~collaboration), et~al., Search for majorana neutrinos with the complete {KamLAND-Zen} dataset, Phys. Rev. Lett. 135~(26) (2025) 262501.
\newblock \href {https://doi.org/10.1103/jkf6-48j8} {\path{doi:10.1103/jkf6-48j8}}.

\bibitem{PandaX-4T:2025jel}
L.~Luo ({PandaX}~collaboration), et~al., Search for double beta decay of \ensuremath{^{136}}{Xe} to the \ensuremath{{0}_1^{+}} excited state of \ensuremath{^{136}}{Ba} with {PandaX-4T}, JHEP 05 (2025) 089.
\newblock \href {https://doi.org/10.1007/JHEP05(2025)089} {\path{doi:10.1007/JHEP05(2025)089}}.

\bibitem{nEXO:2021ujk}
G.~Adhikari ({nEXO}~collaboration), et~al., {nEXO: neutrinoless double beta decay search beyond 10$^{28}$ year half-life sensitivity}, J. Phys. G 49~(1) (2022) 015104.
\newblock \href {https://doi.org/10.1088/1361-6471/ac3631} {\path{doi:10.1088/1361-6471/ac3631}}.

\bibitem{NEXT:2020amj}
C.~Adams ({NEXT}~collaboration), et~al., Sensitivity of a tonne-scale next detector for neutrinoless double beta decay searches, JHEP 2021~(08) (2021) 164.
\newblock \href {https://doi.org/10.1007/JHEP08(2021)164} {\path{doi:10.1007/JHEP08(2021)164}}.

\bibitem{XLZD:2024pdv}
J.~Aalbers~({XLZD collaboration}), et~al., Neutrinoless double beta decay sensitivity of the {XLZD} rare event observatory, J. Phys. G 52~(4) (2025) 045102.
\newblock \href {https://doi.org/10.1088/1361-6471/adb900} {\path{doi:10.1088/1361-6471/adb900}}.

\bibitem{nudat3}
NuDat3.0, \href{http://www.nndc.bnl.gov/nudat3/}{nndc.bnl.gov/nudat3/}.

\bibitem{Akerib2020}
D.~S. Akerib~({LUX collaboration}), et~al., Search for two neutrino double electron capture of ${}^{124}\mathrm{Xe}$ and ${}^{126}\mathrm{Xe}$ in the full exposure of the {{LUX}} detector, J. Phys. G 47~(10) (2020) 105105.
\newblock \href {https://doi.org/10.1088/1361-6471/ab9c2d} {\path{doi:10.1088/1361-6471/ab9c2d}}.

\bibitem{XMASScoll18}
K.~Abe~({XMASS collaboration}), et~al., Improved search for two-neutrino double electron capture on ${}^{124}\mathrm{Xe}$ and ${}^{126}\mathrm{Xe}$ using particle identification in {XMASS-I}, Prog. Theo. Exp. Phys. 2018~(5) (2018) 053D03.
\newblock \href {https://doi.org/10.1093/ptep/pty053} {\path{doi:10.1093/ptep/pty053}}.

\bibitem{EXO2002026}
S.~Al~Kharusi~({EXO-200 collaboration}), et~al., Search for double beta decays of ${}^{134}\mathrm{Xe}$ with {{EXO-200 Phase II}}, Phys. Rev. Lett. 136~(23) (2026) 232502.
\newblock \href {https://doi.org/10.1103/vpny-23wq} {\path{doi:10.1103/vpny-23wq}}.

\bibitem{Yan24}
X.~Yan~({PandaX collaboration}), et~al., Searching for two-neutrino and neutrinoless double beta decay of $^{134}\mathrm{Xe}$ with the {PandaX-4T} experiment, Phys. Rev. Lett. 132 (2024) 152502.
\newblock \href {https://doi.org/10.1103/PhysRevLett.132.152502} {\path{doi:10.1103/PhysRevLett.132.152502}}.

\bibitem{Jokiniemi:2022yfr}
L.~Jokiniemi, B.~Romeo, C.~Brase, J.~Kotila, P.~Soriano, A.~Schwenk, J.~Men{\'e}ndez, Two-neutrino {\ensuremath{\beta}}{\ensuremath{\beta}} decay of \ensuremath{^{136}}{Xe} to the first excited \ensuremath{0^+} state in \ensuremath{^{136}}{Ba}, Phys. Lett. B 838 (2023) 137689.
\newblock \href {https://doi.org/10.1016/j.physletb.2023.137689} {\path{doi:10.1016/j.physletb.2023.137689}}.

\bibitem{Vogel:1986nj}
P.~Vogel, M.~R. Zirnbauer, Suppression of the two neutrino double beta decay by nuclear structure effects, Phys. Rev. Lett. 57 (1986) 3148--3151.
\newblock \href {https://doi.org/10.1103/PhysRevLett.57.3148} {\path{doi:10.1103/PhysRevLett.57.3148}}.

\bibitem{Engel:1988au}
J.~Engel, P.~Vogel, M.~R. Zirnbauer, Nuclear structure effects in double beta decay, Phys. Rev. C 37 (1988) 731--746.
\newblock \href {https://doi.org/10.1103/PhysRevC.37.731} {\path{doi:10.1103/PhysRevC.37.731}}.

\bibitem{Ejiri:2019ezh}
H.~Ejiri, J.~Suhonen, K.~Zuber, Neutrino{\textendash}nuclear responses for astro-neutrinos, single beta decays and double beta decays, Phys. Rept. 797 (2019) 1--102.
\newblock \href {https://doi.org/10.1016/j.physrep.2018.12.001} {\path{doi:10.1016/j.physrep.2018.12.001}}.

\bibitem{Caurier:1990dc}
E.~Caurier, A.~P. Zuker, A.~Poves, A full \ensuremath{0\hbar\omega} description of the \ensuremath{2\nu\beta\beta} decay of \ensuremath{^{48}}{Ca}, Phys. Lett. B 252 (1990) 13--17.
\newblock \href {https://doi.org/10.1016/0370-2693(90)91071-I} {\path{doi:10.1016/0370-2693(90)91071-I}}.

\bibitem{Caurier12}
E.~Caurier, F.~Nowacki, A.~Poves, Shell model description of the $\beta\beta$ decay of $^{136}\mathrm{Xe}$, Phys. Lett. B 711 (2012) 62.
\newblock \href {https://doi.org/10.1016/j.physletb.2012.03.076} {\path{doi:10.1016/j.physletb.2012.03.076}}.

\bibitem{Horoi13}
M.~Horoi, B.~A. Brown, Shell-model analysis of the $^{136}\mathrm{Xe}$ double beta decay nuclear matrix elements, Phys. Rev. Lett. 110 (2013) 222502.
\newblock \href {https://doi.org/10.1103/PhysRevLett.110.222502} {\path{doi:10.1103/PhysRevLett.110.222502}}.

\bibitem{Coraggio:2023eep}
L.~Coraggio, N.~Itaco, G.~De~Gregorio, A.~Gargano, Z.~H. Cheng, Y.~Z. Ma, F.~R. Xu, M.~Viviani, The renormalization of the shell-model {Gamow-Teller} operator starting from effective field theory for nuclear systems, Phys. Rev. C 109~(1) (2024) 014301.
\newblock \href {https://doi.org/10.1103/PhysRevC.109.014301} {\path{doi:10.1103/PhysRevC.109.014301}}.

\bibitem{Barea:2015kwa}
J.~Barea, J.~Kotila, F.~Iachello, $0\nu\beta\beta$ and $2\nu\beta\beta$ nuclear matrix elements in the interacting boson model with isospin restoration, Phys. Rev. C 91~(3) (2015) 034304.
\newblock \href {https://doi.org/10.1103/PhysRevC.91.034304} {\path{doi:10.1103/PhysRevC.91.034304}}.

\bibitem{CoelloPerez_2025}
E.~A. Coello~P\'erez, T.~Papenbrock, Effective field theories for collective excitations of atomic nuclei, J. Phys. G. 52~(3) (2025) 033001.
\newblock \href {https://doi.org/10.1088/1361-6471/adb50b} {\path{doi:10.1088/1361-6471/adb50b}}.

\bibitem{CoelloPerez_2018}
E.~A. Coello~P\'erez, J.~Men\'endez, A.~Schwenk, Gamow-{T}eller and double-$\ensuremath{\beta}$ decays of heavy nuclei within an effective theory, Phys. Rev. C 98 (2018) 045501.
\newblock \href {https://doi.org/10.1103/PhysRevC.98.045501} {\path{doi:10.1103/PhysRevC.98.045501}}.

\bibitem{Delion:2017bie}
D.~S. Delion, J.~Suhonen, Two-neutrino {\ensuremath{\beta}}{\ensuremath{\beta}} decays and low-lying {Gamow-Teller} {\ensuremath{\beta}}{\ensuremath{^-}} strength functions in the mass range {A}=70{\textendash}176, Phys. Rev. C 95~(3) (2017) 034330.
\newblock \href {https://doi.org/10.1103/PhysRevC.95.034330} {\path{doi:10.1103/PhysRevC.95.034330}}.

\bibitem{Raduta:2005be}
A.~A. Raduta, C.~M. Raduta, A.~Escuderos, Description of the two neutrino double beta decay in deformed nuclei with projected spherical single particle basis, Phys. Rev. C 71 (2005) 024307.
\newblock \href {https://doi.org/10.1103/PhysRevC.71.024307} {\path{doi:10.1103/PhysRevC.71.024307}}.

\bibitem{Singh:2007jh}
S.~Singh, R.~Chandra, P.~K. Rath, P.~K. Raina, J.~G. Hirsch, Nuclear deformation and the two neutrino double-beta decay in \ensuremath{^{124,126}}{Xe}, \ensuremath{^{128,130}}{Te}, \ensuremath{^{130,132}}{Ba} and \ensuremath{^{150}}{Nd} isotopes, Eur. Phys. J. A 33 (2007) 375--388.
\newblock \href {https://doi.org/10.1140/epja/i2007-10481-7} {\path{doi:10.1140/epja/i2007-10481-7}}.

\bibitem{Shukla07}
A.~Shukla, P.~K. Raina, P.~K. Rath, Study of two neutrino $\ensuremath{\beta^+\beta^+}/\ensuremath{\beta^+}$ $\mathrm{EC}/\mathrm{ECEC}$ decay of $^{124,126}\mathrm{Xe}$ and $^{130,132}\mathrm{Ba}$ for \ensuremath{0^+\rightarrow0^+} transition in {PHFB} model, J. Phys. G. 34~(3) (2007) 549.
\newblock \href {https://doi.org/10.1088/0954-3899/34/3/013} {\path{doi:10.1088/0954-3899/34/3/013}}.

\bibitem{Gysbers2019a}
P.~Gysbers, G.~Hagen, J.~D. Holt, G.~R. Jansen, T.~D. Morris, P.~Navrátil, T.~Papenbrock, S.~Quaglioni, A.~Schwenk, S.~R. Stroberg, K.~A. Wendt, Discrepancy between experimental and theoretical $\beta$-decay rates resolved from first principles, Nature Phys. 15~(5) (2019) 428--431.
\newblock \href {https://doi.org/10.1038/s41567-019-0450-7} {\path{doi:10.1038/s41567-019-0450-7}}.

\bibitem{Stroberg:2021guc}
S.~R. Stroberg, Beta decay in medium-mass nuclei with the in-medium similarity renormalization group, Particles 4~(4) (2021) 521--535.
\newblock \href {https://doi.org/10.3390/particles4040038} {\path{doi:10.3390/particles4040038}}.

\bibitem{Li:2025exk}
Z.~Li, T.~Miyagi, A.~Schwenk, Ab initio calculations of {\ensuremath{\beta}}-decay half-lives for {N}=50 neutron-rich nuclei, Phys. Rev. Lett. 136~(18) (2026) 182501.
\newblock \href {https://doi.org/10.1103/xjv9-t6sn} {\path{doi:10.1103/xjv9-t6sn}}.

\bibitem{Novario2021}
S.~Novario, P.~Gysbers, J.~Engel, G.~Hagen, G.~R. Jansen, T.~D. Morris, P.~Navr\'atil, T.~Papenbrock, S.~Quaglioni, Coupled-cluster calculations of neutrinoless double-$\ensuremath{\beta}$ decay in $^{48}\mathrm{Ca}$, Phys. Rev. Lett. 126 (2021) 182502.
\newblock \href {https://doi.org/10.1103/PhysRevLett.126.182502} {\path{doi:10.1103/PhysRevLett.126.182502}}.

\bibitem{Li2026}
Z.~Li, L.~Jokiniemi, A.~Schwenk, {Ab initio calculations of two-neutrino and neutrinoless double-$\beta$ decay of $^{48}$Ca and related Gamow-Teller strength distributions} (2026).
\newblock \href {http://arxiv.org/abs/2607.11733} {\path{arXiv:2607.11733}}.

\bibitem{LUXZEPLIN21}
D.~S. Akerib~({LUX-ZEPLIN collaboration}), et~al., Projected sensitivity of the {LUX-ZEPLIN} experiment to the two-neutrino and neutrinoless double $\ensuremath{\beta}$ decays of $^{134}\mathrm{Xe}$, Phys. Rev. C 104 (2021) 065501.
\newblock \href {https://doi.org/10.1103/PhysRevC.104.065501} {\path{doi:10.1103/PhysRevC.104.065501}}.

\bibitem{Simkovic:2018rdz}
F.~\v{S}imkovic, R.~Dvornick\'y, D.~Stef\'anik, A.~Faessler, Improved description of the $2\nu\beta\beta$-decay and a possibility to determine the effective axial-vector coupling constant, Phys. Rev. C 97~(3) (2018) 034315.
\newblock \href {https://doi.org/10.1103/PhysRevC.97.034315} {\path{doi:10.1103/PhysRevC.97.034315}}.

\bibitem{Morabit:2024sms}
S.~e. Morabit, R.~Bouabid, V.~Cirigliano, J.~de~Vries, L.~Gr{\'a}f, E.~Mereghetti, 2{\ensuremath{\nu}}{\ensuremath{\beta}}{\ensuremath{\beta}} spectrum in chiral effective field theory, JHEP 06 (2025) 082.
\newblock \href {https://doi.org/10.1007/JHEP06(2025)082} {\path{doi:10.1007/JHEP06(2025)082}}.

\bibitem{Castillo2025}
D.~Castillo, D.~Frycz, B.~Benavente, J.~Menéndez, Two-neutrino $\beta\beta$ decay to excited states at next-to-leading order, Phys. Lett. B 875 (2026) 140306.
\newblock \href {https://doi.org/10.1016/j.physletb.2026.140306} {\path{doi:10.1016/j.physletb.2026.140306}}.

\bibitem{Kotila2012}
J.~Kotila, F.~Iachello, Phase-space factors for double-$\beta$ decay, Phys. Rev. C 85 (2012) 034316.
\newblock \href {https://doi.org/10.1103/PhysRevC.85.034316} {\path{doi:10.1103/PhysRevC.85.034316}}.

\bibitem{Kotila:2013gea}
J.~Kotila, F.~Iachello, Phase space factors for $\beta^+\beta^+$ decay and competing modes of double-$\beta$ decay, Phys. Rev. C 87 (2013) 024313.
\newblock \href {https://doi.org/10.1103/PhysRevC.87.024313} {\path{doi:10.1103/PhysRevC.87.024313}}.

\bibitem{Faessler1998}
A.~Faessler, F.~{\v{S}}imkovic, Double beta decay, J. Phys. G 24~(12) (1998) 2139--2178.
\newblock \href {https://doi.org/10.1088/0954-3899/24/12/001} {\path{doi:10.1088/0954-3899/24/12/001}}.

\bibitem{Elliott2002}
S.~R. Elliott, P.~Vogel, Double beta decay, Annu. Rev. Nucl. Part. Sci. 52~(1) (2002) 115--151.
\newblock \href {https://doi.org/10.1146/annurev.nucl.52.050102.090641} {\path{doi:10.1146/annurev.nucl.52.050102.090641}}.

\bibitem{Suhonen2007}
J.~Suhonen, From Nucleons to Nucleus: Concepts of Microscopic Nuclear Theory, {Springer-Verlag}, {Berlin Heidelberg}, 2007.

\bibitem{Simkovic2013}
F.~{\v{S}}imkovic, V.~Rodin, A.~Faessler, P.~Vogel, $\ensuremath{0\nu\beta\beta}$ and $\ensuremath{2\nu\beta\beta}$ nuclear matrix elements, quasiparticle random-phase approximation, and isospin symmetry restoration, Phys. Rev. C 87 (2013) 045501.
\newblock \href {https://doi.org/10.1103/PhysRevC.87.045501} {\path{doi:10.1103/PhysRevC.87.045501}}.

\bibitem{Jokiniemi2018}
L.~Jokiniemi, H.~Ejiri, D.~Frekers, J.~Suhonen, Neutrinoless $\beta\beta$ nuclear matrix elements using isovector spin-dipole $\textit{J}^\pi=2^-$ data, Phys. Rev. C 98 (2018) 024608.
\newblock \href {https://doi.org/10.1103/PhysRevC.98.024608} {\path{doi:10.1103/PhysRevC.98.024608}}.

\bibitem{Jokiniemi2021}
L.~Jokiniemi, P.~Soriano, J.~Men\'{e}ndez, Impact of the leading-order short-range nuclear matrix element on the neutrinoless double-beta decay of medium-mass and heavy nuclei, Phys. Lett. B 823 (2021) 136720.
\newblock \href {https://doi.org/10.1016/j.physletb.2021.136720} {\path{doi:10.1016/j.physletb.2021.136720}}.

\bibitem{Holinde1981}
K.~Holinde, Two-nucleon forces and nuclear matter, Phys. Rep. 68 (1981) 121.
\newblock \href {https://doi.org/10.1016/0370-1573(81)90188-5} {\path{doi:10.1016/0370-1573(81)90188-5}}.

\bibitem{MPinedo}
E.~Caurier, G.~Martinez-Pinedo, F.~Nowacki, A.~Poves, A.~P. Zuker, The shell model as unified view of nuclear structure, Rev. Mod. Phys. 77 (2005) 427.
\newblock \href {https://doi.org/10.1103/RevModPhys.77.427} {\path{doi:10.1103/RevModPhys.77.427}}.

\bibitem{Brown01}
B.~A. Brown, The nuclear shell model towards the drip lines, Prog. Part. Nucl. Phys. 47 (2001) 517.
\newblock \href {https://doi.org/10.1016/S0146-6410(01)00159-4} {\path{doi:10.1016/S0146-6410(01)00159-4}}.

\bibitem{Otsuka19}
T.~Otsuka, A.~Gade, O.~Sorlin, T.~Suzuki, Y.~Utsuno, Evolution of shell structure in exotic nuclei, Rev. Mod. Phys. 92 (2020) 015002.
\newblock \href {https://doi.org/10.1103/RevModPhys.92.015002} {\path{doi:10.1103/RevModPhys.92.015002}}.

\bibitem{Wildenthal85}
B.~H. Wildenthal, W.~Chung, B.~A. Brown, Experimental and theoretical {G}amow-{T}eller beta-decay observables for the sd-shell nuclei, At. Data Nucl. Data Tables 33 (1985) 347--404.
\newblock \href {https://doi.org/10.1016/0092-640X(85)90009-9} {\path{doi:10.1016/0092-640X(85)90009-9}}.

\bibitem{Chou93}
W.~T. Chou, E.~K. Warburton, B.~A. Brown, Gamow-{T}eller beta-decay rates for {A} $\leq$ 18 nuclei, Phys. Rev. C 47 (1993) 163.
\newblock \href {https://doi.org/10.1103/PhysRevC.47.163} {\path{doi:10.1103/PhysRevC.47.163}}.

\bibitem{Yoshida18}
S.~Yoshida, Y.~Utsuno, N.~Shimizu, T.~Otsuka, Systematic shell-model study of $\ensuremath{\beta}$-decay properties and {G}amow-{T}eller strength distributions in $\mathrm{A}\ensuremath{\approx}40$ neutron-rich nuclei, Phys. Rev. C 97 (2018) 054321.
\newblock \href {https://doi.org/10.1103/PhysRevC.97.054321} {\path{doi:10.1103/PhysRevC.97.054321}}.

\bibitem{Coraggio20}
L.~Coraggio, A.~Gargano, N.~Itaco, R.~Mancino, F.~Nowacki, Calculation of the neutrinoless double-$\beta$ decay matrix element within the realistic shell model, Phys. Rev. C 101 (2020) 044315.
\newblock \href {https://doi.org/10.1103/PhysRevC.101.044315} {\path{doi:10.1103/PhysRevC.101.044315}}.

\bibitem{Nitescu:2024ppf}
O.~Ni\c{t}escu, S.~Ghinescu, V.-A. Sevestrean, M.~Horoi, F.~\v{S}imkovic, S.~Stoica, Theoretical analysis and predictions for the two-neutrino double electron capture of $^{124}\mathrm{Xe}$, J. Phys. G 51~(12) (2024) 125103.
\newblock \href {https://doi.org/10.1088/1361-6471/ad8767} {\path{doi:10.1088/1361-6471/ad8767}}.

\bibitem{Caurier:2010az}
E.~Caurier, F.~Nowacki, A.~Poves, K.~Sieja, Collectivity in the light xenon isotopes: A shell model study, Phys. Rev. C 82 (2010) 064304.
\newblock \href {https://doi.org/10.1103/PhysRevC.82.064304} {\path{doi:10.1103/PhysRevC.82.064304}}.

\bibitem{QiQX}
C.~Qi, Z.~X. Xu, Monopole-optimized effective interaction for tin isotopes, Phys. Rev. C 86 (2012) 044323.
\newblock \href {https://doi.org/10.1103/PhysRevC.86.044323} {\path{doi:10.1103/PhysRevC.86.044323}}.

\bibitem{Rebeiro:2023kvs}
B.~M. Rebeiro, S.~Triambak, P.~E. Garrett, G.~C. Ball, B.~A. Brown, J.~Men\'endez, B.~Romeo, P.~Adsley, B.~G. Lenardo, R.~Lindsay, V.~Bildstein, C.~Burbadge, R.~Coleman, A.~Diaz~Varela, R.~Dubey, T.~Faestermann, R.~Hertenberger, M.~Kamil, K.~G. Leach, C.~Natzke, J.~C. Nzobadila~Ondze, A.~Radich, E.~Rand, H.-F. Wirth, $^{138}\mathrm{Ba}$(d,\ensuremath{\alpha}) study of states in $^{136}\mathrm{Cs}$: Implications for new physics searches with xenon detectors, Phys. Rev. Lett. 131~(5) (2023) 052501.
\newblock \href {https://doi.org/10.1103/PhysRevLett.131.052501} {\path{doi:10.1103/PhysRevLett.131.052501}}.

\bibitem{Horoi16}
M.~Horoi, A.~Neacsu, Shell model predictions for $^{124}\mathrm{Sn}$ double-$\ensuremath{\beta}$ decay, Phys. Rev. C 93 (2016) 024308.
\newblock \href {https://doi.org/10.1103/PhysRevC.93.024308} {\path{doi:10.1103/PhysRevC.93.024308}}.

\bibitem{ARIMA1977205}
A.~Arima, T.~Otsuka, F.~Iachello, I.~Talmi, Collective nuclear states as symmetric couplings of proton and neutron excitations, Phys. Lett. B 66 (1977) 205.
\newblock \href {https://doi.org/10.1016/0370-2693(77)90860-7} {\path{doi:10.1016/0370-2693(77)90860-7}}.

\bibitem{iac87}
F.~Iachello, A.~Arima, The Interacting Boson Model, Cambridge University Press, 1987.

\bibitem{OTSUKA19781}
T.~Otsuka, A.~Arima, F.~Iachello, Nuclear shell model and interacting bosons, Nucl. Phys. A 309 (1978) 1.
\newblock \href {https://doi.org/10.1016/0375-9474(78)90532-8} {\path{doi:10.1016/0375-9474(78)90532-8}}.

\bibitem{Barea:2009zza}
J.~Barea, F.~Iachello, Neutrinoless double-beta decay in the microscopic interacting boson model, Phys. Rev. C 79 (2009) 044301.
\newblock \href {https://doi.org/10.1103/PhysRevC.79.044301} {\path{doi:10.1103/PhysRevC.79.044301}}.

\bibitem{GADE2000268}
A.~Gade, I.~Wiedenhöver, J.~Gableske, A.~Gelberg, H.~Meise, N.~Pietralla, P.~{von Brentano}, Proton–neutron structure of low lying collective quadrupole excitations in ${}^{126}\mathrm{Xe}$, Nucl. Phys. A 665~(3) (2000) 268--284.
\newblock \href {https://doi.org/10.1016/S0375-9474(99)00387-5} {\path{doi:10.1016/S0375-9474(99)00387-5}}.

\bibitem{PUDDU1980109}
G.~Puddu, O.~Scholten, T.~Otsuka, Collective quadrupole states of $\mathrm{Xe,~Ba~and~Ce}$ in the interacting boson model, Nucl. Phys. A 348 (1980) 109.
\newblock \href {https://doi.org/10.1016/0375-9474(80)90548-5} {\path{doi:10.1016/0375-9474(80)90548-5}}.

\bibitem{Kotila:2016pib}
J.~Kotila, J.~Barea, Occupation probabilities of single particle levels using the microscopic interacting boson model: Application to some nuclei of interest in neutrinoless double-$\beta$ decay, Phys. Rev. C 94 (2016) 034320.
\newblock \href {https://doi.org/10.1103/PhysRevC.94.034320} {\path{doi:10.1103/PhysRevC.94.034320}}.

\bibitem{Abad1984a}
J.~Abad, A.~Morales, R.~Nu{\~n}ez-Lagos, A.~F. Pacheco, An estimation of the rates of (two-neutrino) double beta decay and related processes, Ann. Fis. Ser. A 80 (1984) 9.

\bibitem{Abad1984b}
J.~Abad, A.~Morales, R.~Nu{\~n}ez-Lagos, A.~F. Pacheco, An estimation of the rates of (two-neutrino) double beta decay and related processes, J Phys. Colloques 45 (1984) C3--147.
\newblock \href {https://doi.org/10.1051/jphyscol:1984328} {\path{doi:10.1051/jphyscol:1984328}}.

\bibitem{Domin2005}
P.~Domin, S.~Kovalenko, F.~{\v{S}}imkovic, S.~V. Semenov, Neutrino accompanied $\beta^\pm\beta^\pm$, $\beta^+/\mathrm{EC}$ and $\mathrm{EC}/\mathrm{EC}$ processes within single state dominance hypothesis, Nucl. Phys. A 753 (2005) 337.
\newblock \href {https://doi.org/10.1016/j.nuclphysa.2005.03.003} {\path{doi:10.1016/j.nuclphysa.2005.03.003}}.

\bibitem{PhysRevC.87.014315}
J.~Barea, J.~Kotila, F.~Iachello, Nuclear matrix elements for double-$\ensuremath{\beta}$ decay, Phys. Rev. C 87 (2013) 014315.
\newblock \href {https://doi.org/10.1103/PhysRevC.87.014315} {\path{doi:10.1103/PhysRevC.87.014315}}.

\bibitem{logft_code_LogFT}
LogFT, \href{http://www.nndc.bnl.gov/logft/}{nndc.bnl.gov/logft/}.

\bibitem{logft_code_BetaShape}
BetaShape, \href{http://www.lnhb.fr/home/rd-activities/spectrum-processing-software/}{lnhb.fr/home/rd-activities/spectrum-processing-software/}.

\bibitem{Mougeot_2019}
X.~Mougeot, Towards high-precision calculation of electron capture decays, Appl. Radiat. Isot. 154 (2019) 108884.
\newblock \href {https://doi.org/10.1016/j.apradiso.2019.108884} {\path{doi:10.1016/j.apradiso.2019.108884}}.

\bibitem{Mougeot_2023}
X.~Mougeot, Atomic exchange correction in forbidden unique beta transitions, Appl. Radiat. Isot. 201 (2023) 111018.
\newblock \href {https://doi.org/10.1016/j.apradiso.2023.111018} {\path{doi:10.1016/j.apradiso.2023.111018}}.

\bibitem{Nitescu2025}
O.~Ni\ifmmode~\mbox{\c{t}}\else \c{t}\fi{}escu, F.~\ifmmode~\check{S}\else \v{S}\fi{}imkovic, Semi-empirical formula for two-neutrino double-$\ensuremath{\beta}$ decay, Phys. Rev. C 111 (2025) 024307.
\newblock \href {https://doi.org/10.1103/PhysRevC.111.024307} {\path{doi:10.1103/PhysRevC.111.024307}}.

\bibitem{KamLAND-Zen2019}
A.~Gando~({KamLAND-Zen collaboration}), et~al., Precision measurement of the \ensuremath{^{136}}{Xe} two-neutrino \ensuremath{\beta\beta} spectrum in {KamLAND-Zen} and its impact on the quenching of nuclear matrix elements, Phys. Rev. Lett. 122~(19) (2019) 192501.
\newblock \href {https://doi.org/10.1103/PhysRevLett.122.192501} {\path{doi:10.1103/PhysRevLett.122.192501}}.

\bibitem{Wang_2021}
M.~Wang, W.~Huang, F.~Kondev, G.~Audi, S.~Naimi, The {AME} 2020 atomic mass evaluation (ii). tables, graphs and references*, Chin. Phys. C 45~(3) (2021) 030003.
\newblock \href {https://doi.org/10.1088/1674-1137/abddaf} {\path{doi:10.1088/1674-1137/abddaf}}.

\bibitem{ensdf}
ENSDF, \href{http://www.nndc.bnl.gov/ensdf/}{nndc.bnl.gov/ensdf/}.

\bibitem{Nitescu24}
O.~Niţescu, S.~Ghinescu, V.~A. Sevestrean, M.~Horoi, F.~Šimkovic, S.~Stoica, Theoretical analysis and predictions for the two-neutrino double electron capture of \ensuremath{^{124}}{Xe}, {J. Phys. G.} 51~(12) (2024) 125103.
\newblock \href {https://doi.org/10.1088/1361-6471/ad8767} {\path{doi:10.1088/1361-6471/ad8767}}.

\bibitem{Puppe2011}
P.~Puppe, D.~Frekers, T.~Adachi, H.~Akimune, N.~Aoi, B.~Bilgier, H.~Ejiri, H.~Fujita, Y.~Fujita, M.~Fujiwara, E.~Ganio\ifmmode~\breve{g}\else \u{g}\fi{}lu, M.~N. Harakeh, K.~Hatanaka, M.~Holl, H.~C. Kozer, J.~Lee, A.~Lennarz, H.~Matsubara, K.~Miki, S.~E.~A. Orrigo, T.~Suzuki, A.~Tamii, J.~H. Thies, High-resolution (${}^{3}\mathrm{He}$,$t$) reaction on the double-$\ensuremath{\beta}$ decaying nucleus ${}^{136}\mathrm{Xe}$, Phys. Rev. C 84 (2011) 051305.
\newblock \href {https://doi.org/10.1103/PhysRevC.84.051305} {\path{doi:10.1103/PhysRevC.84.051305}}.

\end{thebibliography}

\end{document}